
\hoffset -0.25cm
\voffset -0.0cm
\magnification=\magstep1
\baselineskip=15pt
\hsize=16truecm
\vsize=23truecm
\font\fontA=cmr10
\font\fontB=cmbx10
\font\fontI=cmti10
\newcount\EQnum 
\newcount\FGnum 
\newcount\SCnum 
\newcount\TBnum 
\newcount\RFnum 
%
\global\SCnum=0
\global\EQnum=0
\global\FGnum=0
\global\RFnum=0
\global\TBnum=0
%
\newcount\SCnumA
\newcount\SCnumB
\newcount\SCnumC
\newcount\SCnumD
\newcount\SCnumE
\newcount\SCnumF
\newcount\SCnumG
\newcount\SCnumH
\newcount\SCnumI
\newcount\SCnumJ
\newcount\SCnumK
\newcount\SCnumL
\newcount\SCnumM
\newcount\SCnumN
\newcount\SCnumO
\newcount\SCnumP
\newcount\SCnumQ
\newcount\SCnumR
\newcount\SCnumS
\newcount\SCnumT
\newcount\SCnumU
\newcount\SCnumV
\newcount\SCnumW
\newcount\SCnumX
\newcount\SCnumY
\newcount\SCnumZ
%
\newcount\EQnumA
\newcount\EQnumB
\newcount\EQnumC
\newcount\EQnumD
\newcount\EQnumE
\newcount\EQnumF
\newcount\EQnumG
\newcount\EQnumH
\newcount\EQnumI
\newcount\EQnumJ
\newcount\EQnumK
\newcount\EQnumL
\newcount\EQnumM
\newcount\EQnumN
\newcount\EQnumO
\newcount\EQnumP
\newcount\EQnumQ
\newcount\EQnumR
\newcount\EQnumS
\newcount\EQnumT
\newcount\EQnumU
\newcount\EQnumV
\newcount\EQnumW
\newcount\EQnumX
\newcount\EQnumY
\newcount\EQnumZ
%
\newcount\FGnumA
\newcount\FGnumB
\newcount\FGnumC
\newcount\FGnumD
\newcount\FGnumE
\newcount\FGnumF
\newcount\FGnumG
\newcount\FGnumH
\newcount\FGnumI
\newcount\FGnumJ
\newcount\FGnumK
\newcount\FGnumL
\newcount\FGnumM
\newcount\FGnumN
\newcount\FGnumO
\newcount\FGnumP
\newcount\FGnumQ
\newcount\FGnumR
\newcount\FGnumS
\newcount\FGnumT
\newcount\FGnumU
\newcount\FGnumV
\newcount\FGnumW
\newcount\FGnumX
\newcount\FGnumY
\newcount\FGnumZ
%
\newcount\RFnumA
\newcount\RFnumB
\newcount\RFnumC
\newcount\RFnumD
\newcount\RFnumE
\newcount\RFnumF
\newcount\RFnumG
\newcount\RFnumH
\newcount\RFnumI
\newcount\RFnumJ
\newcount\RFnumK
\newcount\RFnumL
\newcount\RFnumM
\newcount\RFnumN
\newcount\RFnumO
\newcount\RFnumP
\newcount\RFnumQ
\newcount\RFnumR
\newcount\RFnumS
\newcount\RFnumT
\newcount\RFnumU
\newcount\RFnumV
\newcount\RFnumW
\newcount\RFnumX
\newcount\RFnumY
\newcount\RFnumZ
%
\newcount\RFnumA
\newcount\TBnumB
\newcount\TBnumC
\newcount\TBnumD
\newcount\TBnumE
\newcount\TBnumF
\newcount\TBnumG
\newcount\TBnumH
\newcount\TBnumI
\newcount\TBnumJ
\newcount\TBnumK
\newcount\TBnumL
\newcount\TBnumM
\newcount\TBnumN
\newcount\TBnumO
\newcount\TBnumP
\newcount\TBnumQ
\newcount\TBnumR
\newcount\TBnumS
\newcount\TBnumT
\newcount\TBnumU
\newcount\TBnumV
\newcount\TBnumW
\newcount\TBnumX
\newcount\TBnumY
\newcount\TBnumZ
%
\def\ie{{\fontI i.e.}}

\def\RF{$^{\global\advance\RFnum by 1
           \number\RFnum)}\ $}
\def\RFRF{$^{\global\advance\RFnum by 1
             \number\RFnum,
             \global\advance\RFnum by 1
             \number\RFnum)}\ $}
\def\RFRFRF{$^{\global\advance\RFnum by 1
               \number\RFnum-
               \global\advance\RFnum by 2
               \number\RFnum)}\ $}
\def\RFRFRFRF{$^{\global\advance\RFnum by 1
                 \number\RFnum-
                 \global\advance\RFnum by 3
                 \number\RFnum)}\ $}
\def\RFRFRFRFRF{$^{\global\advance\RFnum by 1
                   \number\RFnum-
                   \global\advance\RFnum by 4
                   \number\RFnum)}\ $}
\def\RFRFRFRFRFRF{$^{\global\advance\RFnum by 1
                     \number\RFnum-
                     \global\advance\RFnum by 5
                     \number\RFnum)}\ $}
\def\RFRFRFRFRFRFRF{$^{\global\advance\RFnum by 1
                       \number\RFnum-
                       \global\advance\RFnum by 6
                       \number\RFnum)}\ $}
\def\RFRFRFRFRFRFRFRF{$^{\global\advance\RFnum by 1
                         \number\RFnum-
                         \global\advance\RFnum by 7
                         \number\RFnum)}\ $}
\def\REF{\global\advance\RFnum by 1 \item{\number\RFnum )}}
\def\FIG{\global\advance\FGnum by 1 \item{{\bf Fig.}\ \number\FGnum}}
\def\TAB{\global\advance\TBnum by 1
         \item{{\bf Table} \uppercase\expandafter{\romannumeral\TBnum}}}

%
\def\journal #1#2#3#4{#1 {\bf #2} (#4) #3}
\def\PR{Phys.\ Rev.}
\def\PRB{Phys.\ Rev.\ B}
\def\PRL{Phys.\ Rev.\ Lett.}

\def\IJMP{Int.\ J.\ Mod.\ Phys.}
\def\PLA{Phys.\ Lett.\ A}

\def\JPC{J.\ Phys.\ C}

\def\JPSJ{J.\ Phys.\ Soc.\ Jpn.}

\def\RMP{Rev.\ Mod.\ Phys.}
\def\PTP{Prog.\ Theor.\ Phys.}

%
\hyphenation{Coul-omb}
\hyphenation{pho-non}
\hyphenation{pho-nons}
\hyphenation{Phys-ics}
\hyphenation{phys-ics}
\hyphenation{There-by}
\hyphenation{var-i-a-tion-al}
\hyphenation{anti-ferro-mag-net}
\hyphenation{anti-ferro-mag-nets}
\hyphenation{anti-ferro-mag-netism}
\hyphenation{Gutz-wil-ler}
\def\journal#1#2#3#4{#1 {\bf #2}, #3 (#4)}  



\gdef\refto#1{$^{#1)}$}
\gdef\reftorange#1#2#3{$^{\cite{#1}-\setbox0=\hbox{\cite{#2}}\cite{#3})}$}
\gdef\refis#1{\item{#1)\ }}

\def\references {\vfill\eject {\centerline {\bf References}}\medskip}
\def\endreferences{\vfill\eject}

\catcode`@=11
\newcount\r@fcount \r@fcount=0
\newcount\r@fcurr
\immediate\newwrite\reffile
\newif\ifr@ffile\r@ffilefalse
\def\w@rnwrite#1{\ifr@ffile\immediate\write\reffile{#1}\fi\message{#1}}

\def\writer@f#1>>{}
\def\referencefile{
\r@ffiletrue\immediate\openout\reffile=ref
  \def\writer@f##1>>{\ifr@ffile\immediate\write\reffile%
    {\noexpand\refis{##1} = \csname r@fnum##1\endcsname = %
     \expandafter\expandafter\expandafter\strip@t\expandafter%
     \meaning\csname r@ftext\csname r@fnum##1\endcsname\endcsname}\fi}%
  \def\strip@t##1>>{}}

\def\citeall#1{\xdef#1##1{#1{\noexpand\cite{##1}}}}
\def\cite#1{\each@rg\citer@nge{#1}} 

\def\each@rg#1#2{{\let\thecsname=#1\expandafter\first@rg#2,\end,}}
\def\first@rg#1,{\thecsname{#1}\apply@rg} 
\def\apply@rg#1,{\ifx\end#1\let\next=\relax
\else,\thecsname{#1}\let\next=\apply@rg\fi\next}

\def\citer@nge#1{\citedor@nge#1-\end-} 
\def\citer@ngeat#1\end-{#1}
\def\citedor@nge#1-#2-{\ifx\end#2\r@featspace#1 
  \else\citel@@p{#1}{#2}\citer@ngeat\fi} 
\def\citel@@p#1#2{\ifnum#1>#2{\errmessage{Reference range #1-#2\space is bad.}%
    \errhelp{If you cite a series of references by the notation M-N, then M and
    N must be integers, and N must be greater than or equal to M.}}\else%
 {\count0=#1\count1=#2\advance\count1
by1\relax\expandafter\r@fcite\the\count0,%
  \loop\advance\count0 by1\relax
    \ifnum\count0<\count1,\expandafter\r@fcite\the\count0,%
  \repeat}\fi}

\def\r@featspace#1#2 {\r@fcite#1#2,} 
\def\r@fcite#1,{\ifuncit@d{#1}
    \newr@f{#1}%
    \expandafter\gdef\csname r@ftext\number\r@fcount\endcsname%
                     {\message{Reference #1 to be supplied.}%
                      \writer@f#1>>#1 to be supplied.\par}%
 \fi%
 \csname r@fnum#1\endcsname}
\def\ifuncit@d#1{\expandafter\ifx\csname r@fnum#1\endcsname\relax}%
\def\newr@f#1{\global\advance\r@fcount by1%
    \expandafter\xdef\csname r@fnum#1\endcsname{\number\r@fcount}}

\let\r@fis=\refis   
\def\refis#1#2#3\par{\ifuncit@d{#1}
   \newr@f{#1}%
   \w@rnwrite{Reference #1=\number\r@fcount\space is not cited up to now.}\fi%
  \expandafter\gdef\csname r@ftext\csname r@fnum#1\endcsname\endcsname%
  {\writer@f#1>>#2#3\par}}

\def\ignoreuncited{
   \def\refis##1##2##3\par{\ifuncit@d{##1}%
     \else\expandafter\gdef\csname r@ftext\csname
r@fnum##1\endcsname\endcsname%
     {\writer@f##1>>##2##3\par}\fi}}

\def\r@ferr{\endreferences\errmessage{I was expecting to see
\noexpand\endreferences before now;  I have inserted it here.}}
\let\r@ferences=\references
\def\references{\r@ferences\def\endmode{\r@ferr\par\endgroup}}

\let\endr@ferences=\endreferences
\def\endreferences{\r@fcurr=0
  {\loop\ifnum\r@fcurr<\r@fcount
    \advance\r@fcurr by 1\relax\expandafter\r@fis\expandafter{\number\r@fcurr}%
    \csname r@ftext\number\r@fcurr\endcsname%
  \repeat}\gdef\r@ferr{}\endr@ferences}


\let\r@fend=\endpaper\gdef\endpaper{\ifr@ffile
\immediate\write16{Cross References written on []\jobname.ref.}\fi\r@fend}

\catcode`@=12

\citeall\refto  

\ignoreuncited
\fontB
\bigskip\medskip
\centerline{One-Dimensional $t$-$J$ Model from a Variational Viewpoint}
\bigskip
\bigskip
\centerline{Hisatoshi Yokoyama$^{(1) *}$ and Masao Ogata$^{(2) **}$}
\bigskip
\medskip
\fontI
\centerline{$^{(1)}$Department of Physics, Tohoku University,}
\centerline{Aramaki Aoba, Aoba-ku, Sendai 980-77, JAPAN}
\medskip
\centerline{$^{(2)}$Institute of Physics, College of Arts and Sciences,
University of Tokyo}
\centerline{Komaba, Meguro-ku, Tokyo 153, JAPAN}
\bigskip
\medskip
\fontA
\centerline{(Received \qquad\qquad\qquad\qquad)}
\par
\bigskip
\medskip
\baselineskip=15pt
\beginsection Abstract
\par\noindent
The one-dimensional (1D) $t$-$J$ model is investigated by using
a Gutzwiller-Jastrow-type variation method and the exact
diagonalization of small systems.
Variational expectation values are estimated by the variational
Monte Carlo method with sufficient accuracy.
First, we represent the diagonalization results.
Physical quantities like momentum distribution and some correlation
functions show some behaviors which are not expected in repulsive
models, as the value of $J/t$ increases.
These properties as well as energy are well understood
by introducing intersite correlation factors into wave functions.
The phase transition to a separated phase in large-$J/t$
region can be described by an attractive Jastrow wave function
in quantitative agreement with the exact results.
On the other hand, for the supersymmetric case ($J/t=2$) the
original Gutzwiller wave function becomes an extremely good trial
function for all the range of electron density.
Here a kind of \lq\lq free electron" state is realized,
particularly in the low electron density.
Next, the above wave functions are
compared with the Tomonaga-Luttinger-liquid-type wave
function proposed by Hellberg and Mele.
It is found that the correlation factors in
short distances control bulk quantities like energy and the
magnitude of correlation functions, while the long-range part
of correlataion factors determines the critical behavior of
correlation functions.
Lastly, using these functions, charge and spin susceptibilities
and magnetization curve are estimated, which agree with the
exact results.
It is shown that the Mott transition in 1D $t$-$J$ model is
quite different from the Brinkman-Rice transition.
\par
\bigskip\noindent
PACS numbers: 75.10.Jm, 71.10.+x, 74.65.+n, 74.70.Vy
\par
\vfil\eject
\beginsection \S1. Introduction
\par
The $t$-$J$ model is an important model to study highly correlated
electron systems for its simplicity and close relationship to the
high temperature superconductivity.\refto{Anderson1}
Many properties in one-dimensional (1D) systems have been clarified
extensively by a number of methods:
Bethe-Ansatz solutions, $g$-ology, Tomonaga-Luttinger (TL) liquid
theory, quantum Monte Carlo simulations, exact diagonalization
studies of small clusters, and conformal field theory.
We expect that the study of 1D systems will shed light on
more realistic higher-dimensional systems and that the
comparison of the various methods with
the well-established 1D results enables us to judge the validity of
such methods and approximations.
\par

In contrast to the 1D Hubbard model,\refto{LiebWu}
the Bethe Ansatz solution does not exist in the 1D $t$-$J$ model
except for $J/t=0$ (spinless-fermion case) and $J/t=2$
(supersymmetric case).\refto{tjBA,tjBA2}
In both soluble cases, TL liquid\refto{Haldane} is realized and
the exponents of
long-range behaviors of correlation functions were calculated
exactly by combining the Bethe Ansatz equations and the conformal
field theory.\reftorange{Schulz}{FK,KY1}{KY2}
Also obtained were bulk quantities like spin susceptibility
$\chi_{\rm s}$, charge susceptibility $\chi_{\rm c}$, specific
heat coefficient and effective transport mass, which characterize
metal-insulator (Mott) transitions.\refto{FI}
On the other hand, for general values of $J/t$,
Ogata et al.\refto{Ogata} studied the low-lying energy
spectrum of finite systems to obtain the correlation exponents.
In the phase diagram of $J/t$ and the electron density $n$,
the TL liquid theory holds in the small-$J$ region below
$J_{\rm c}/t=2.5\sim3.5$, depending on $n$.
A phase separation takes place in the larger-$J/t$; there is a
region in which the superconducting correlation is dominant,
between the phase separation and the supersymmetric case.
\par

Meanwhile, as for physical quantities such as momentum distribution
function or spin- and charge-correlation functions,
only the long-range behaviors were clarified in the above
analytic methods; the global features were calculated numerically.
In the limit of $J/t\rightarrow0$, identical with the
large-$U$ limit of the Hubbard model,
the correlation functions were obtained by taking advantage of
the spin-charge separation in the ground state.\refto{OgataShiba}
For the other values of $J/t$, Assaad and W\"urtz\refto{Assaad}
and Imada\refto{Imada} have carried out quantum Monte Carlo
simulations.
Pruschke and Shiba\refto{SP} studied the superconducting
correlation functions by the exact diagonalization.
All these results are consistent with the correlation exponents
obtained by the analytic methods.
\par

Although the ground-state properties in the 1D $t$-$J$ model
have been clarified quite well, it is still important to examine
variational wave functions.
For the explicit form of the wave function will make the
complicated physics easy to grasp.
So far, various kinds of variational states have been proposed
for strongly correlated
systems.\reftorange{Gros}{YS1,Voll1,Voll2,YS2,Hellberg1,YO}{Hellberg2}
The Gutzwiller wave function
(GWF)\refto{Gutzwiller} was studied numerically\refto{Gros,YS1} and
analytically.\refto{Voll1,Voll2}
These studies concluded that the GWF is excellent for the
one-dimensional Heisenberg model, but is unsatisfactory, even
qualitatively, in describing the
properties of the strong coupling Hubbard model or of the
small-$J$ region of the $t$-$J$ model.
For example, the GWF does not reproduce the $2k_{\rm F}$ peak
in the spin correlation function;
in the momentum distribution, it has a strange enhancement
for $k>k_{\rm F}$.
The main reason is that the density correlation is not
sufficiently introduced in the GWF, although the spin correlation
is well incorporated.
These unsatisfactory features are partly remedied by introducing
Jastrow-type intersite correlation factors.\refto{YS2,YO,Hellberg2}
\par

In this paper we show that the ground state properties
of the 1D $t$-$J$ model become more easily accessible
from a variational viewpoint.
First, the behaviors obtained by the
exact diagonalization for $J/t=2$ is described extremely
well by the GWF.
Next, the wave function is improved for other values of $J/t$
by introducing Jastrow-type wave functions with intersite correlation
factors, which are classified into two types: one is the
conventional repulsive and attractive Jastrow-correlation factors
and the other is a long-range Jastrow factor introduced by
Hellberg and Mele.\refto{Hellberg2}
The exponent in the former state is the same as the Fermi liquid,
while the latter state has non-trivial exponents.
It is found that the long-range behavior of the Jastrow factor is
essential for the non-trivial exponent, or non-Fermi liquid behavior.
On the other hand, the variational energy and the global features
of the correlation functions are modified mainly by the
short-range behavior of the Jastrow factor.
We also show that quantities like $\chi_{\rm c}$, $\chi_{\rm s}$
and magnetization curve obtained by the above wave functions are
not only qualitatively but quantitatively consistent with the
exact results.
This aspect is in sharp contrast with the so-called Brinkman-Rice
transition.

The outline of this paper is as follows: in \S2 we discuss the
diagonalization results for various correlation
functions as well as the ground-state energy.
In \S3 the properties of the GWF are studied comparing with those
of the exact calculations for the supersymmetric case.
Section 4 is devoted to the study of the other values of $J/t$
by introducing Fermi-liquid-type Jastrow wave functions.
In \S5 another variational state with an essentially long-range
Jastrow correlation (TL-liquid-type) is examined and compared
with the former type.
In \S6, $\chi_{\rm c}$, $\chi_{\rm s}$ and magnetization curve
are investigated.
Section 7 is assigned to summary.
In Appendices A and B, an analytical approach to the GWF used in \S3
and the behavior of $\chi_{\rm c}$ and $\chi_{\rm s}$ in the
Gutzwiller approximation compared in \S6 are summarized respectively.
A part of the results in this paper was reported before.\refto{YO,YKO}

\par
\fontB\noindent
\beginsection \S2. Ground-State Properties Obtained in Small Clusters
\par
\fontA\smallskip
\par\noindent
We study the one-dimensional (1D) $t$-$J$ model defined as
$$
{\cal H}={\cal H}_t+{\cal H}_J , \eqno(2.1)
$$
$$
\eqalignno{
     {\cal H}_t&=-t\sum_{j\sigma}(c_{j\sigma}^\dagger c_{j+1\sigma}+
           {\rm h.c.}),    &(2.1{\rm a})\cr
     {\cal H}_J&=J\sum_j ({\bf S}_j\cdot {\bf S}_{j+1}-
          {1 \over 4}n_jn_{j+1}),                &(2.1{\rm b})\cr
}$$
in the subspace with no double occupancy with $t,J\geq0$.
Spin operators vanish when they are
applied to empty sites.
Henceforth we take $t$ as the unit of energy.

We use the Lanczos method and the conjugate gradient
method\refto{Nishimori} to
obtain the ground-state wave function in small clusters.
Figures 1(a)-(c) show the momentum distribution function
$$
n(k) = \langle c_{k\sigma}^\dagger c_{k\sigma}\rangle,
\eqno{(2.2)}
$$
and spin- and charge-correlation functions
$$
\eqalignno{
S(k) &= {1 \over N_{\rm a}}\sum_{j,\ell} 4\langle S^z_j S^z_\ell \rangle
e^{ik(r_j-r_\ell)}, \cr
N(k) &= {1 \over N_{\rm a}}\sum_{j,\ell} \{\langle n_j n_\ell \rangle
-\langle n_j\rangle \langle n_\ell\rangle \} e^{ik(r_j-r_\ell)},
&{(2.3)}
}
$$
obtained exactly for various values of $J$.
These results are for the quarter-filled case: $n=N/N_{\rm a} = 0.5$,
with $N$ and $N_{\rm a}$ being the
number of electrons and sites, respectively.
As shown in the 1D $U\rightarrow\infty$ Hubbard model,\refto{OgataShiba}
even the 16-site cluster gives fairly reliable correlation functions.
The thermodynamic limit is estimated easily by interpolating the
data points for the 16-site cluster except for the singularities.
Notice that $S(k)$ and $N(k)$ are also calculated by
Assaad and W\"urtz\refto{Assaad} using quantum
Monte Carlo simulations.
Their results agree with Figs.\ 1(a)-(c).

There are power-law singularities in the momentum distribution $n(k)$
at $k_{\rm F}$ and $3k_{\rm F}$.
The $3k_{\rm F}$ singularity is not so prominent in the small systems
in Fig.\ 1(a).
The power-law singularity at $k_{\rm F}$
is the strongest when $J/t=0$.
With increasing $J/t$, its singularity becomes weaker until $J/t=2$
where $n(k)$ seems to have a jump at $k_{\rm F}$.
In the region $J/t>2$ the power-law singularity shows up again.
This dependence on $J$ is consistent with the behavior of the
numerically obtained correlation exponent $K_\rho$
which increases from $1/2$ as $J$ increases.\refto{Ogata}
Here the exponent $K_\rho$ is defined such that
$\langle s_i^zs_j^z \rangle$ decays as
$\exp\{2ik_{\rm F}r_{ij}\}/r_{ij}^{K_\rho+1}$.
The exponent at $k_{\rm F}$ of $n(k)$ is defined as
$n(k)=n_0-C{\rm sgn}(k-k_{\rm F})|k-k_{\rm F}|^\alpha$.
Since this exponent $\alpha$ is related to $K_\rho$ as
$\alpha=(K_\rho+1/K_\rho-2)/4$,
it starts from ${1\over 8}$ at $J/t=0$, decreases for $0<J/t<2$,
becomes $\alpha=0$ near $J/t=2$ and then begins
to increase for $J/t>2$ where $K_\rho$ becomes larger than 1.
\par

There are $2k_{\rm F}$ singularities in $S(k)$ and $N(k)$.
The TL liquid theory predicts that these singularities
have the same correlation exponents $K_\rho$ and that the
size-dependence is $A - C N_{\rm a}^{-K_\rho}$.
As seen from Fig.\ 1, it is difficult to estimate $K_\rho$ from the
size-dependence of the $2k_{\rm F}$ peak.
Especially the peak in $N(k)$ is much weaker than that in $S(k)$.
We expect the coefficient $C$ of the size-dependence $N_{\rm a}^{-K_\rho}$
is very small for $N(k)$.

The global features of the correlation functions for small values of $J$
resemble the results in the large-$U$ Hubbard model.\refto{OgataShiba}
As $J$ increases they lose this behavior and
near the supersymmetric case ($J/t=2$),
the system behaves similarly to the non-interacting case.
This again corresponds to the fact that the exponent $K_\rho$ becomes 1
(free-electron value)
near $J/t=2$ ($J/t\sim 2.3$ for $n=0.5$).\refto{Ogata}
$N(k)$ and $S(k)$ become almost flat in the region $k>2k_{\rm F}$,
which is the same behavior as in the non-interacting case.
However, note that this global features of correlation functions
are non-trivial even if $K_\rho =1$, since this exponent only
guarantees that the long-range behavior of correlation functions
is the same as in the non-interacting case.
Actually the absolute value in the flat region is quite different
from the non-interacting value.
For the case of 8 electrons in 16 sites, we get
$$
\eqalignno{N(k>2k_{\rm F}) &= 0.312 \sim 0.319,  \cr
           S(k>2k_{\rm F}) &= 0.698 ,
&(2.4)}
$$
while in the non-interacting case, $N(k>2k_{\rm F})=S(k>2k_{\rm F})=n=0.5$.
The sum of these two values is, however,
$$
N(k>2k_{\rm F})+S(k>2k_{\rm F})= 1.001 \sim 1.009,  \eqno{(2.5)}
$$
and this is surprisingly close to the non-interacting value ($2n=1$).
We will discuss this result in the light of the GWF in the following chapter.

Next we estimate the ground-state energy.
The ground state is always singlet and non-degenerate,
if we choose periodic (antiperiodic) boundary conditions
for $N/2=$ odd (even), respectively.\refto{OgataShiba,Ogata}
Under these boundary conditions, the energy converges smoothly
to the thermodynamic limit.
For $n=0.5$ we calculate the ground-state energies in
4, 8, 12, and 16 sites clusters
and fit the results to the formula
$$
E/N_{\rm a}=
\epsilon_\infty + C_1/N_{\rm a}^2 + C_2/N_{\rm a}^4 + C_3/N_{\rm a}^6.
\eqno{(2.6)}
$$
In Fig.\ 2 the fitted values of $\epsilon_\infty$ are shown by a
solid line.
In the region $J/t\geq 3.4$
the energies cannot be fitted to this formula because the
system phase separates in this region and the size-dependence
is different from (2.6).
To check the convergence for $J/t<3.4$, we calculate another series of
singlet energies by using a different boundary conditions, \ie,
antiperiodic ones for $N/2=$ odd and vice versa.\refto{footnote1}
Fitting of the data to the same formula (2.6) gives another estimate of
the ground-state energy, ($\epsilon'_\infty$).
[In this case $C_1>0$, while the former fitting gives $C_1<0$.]
The fitted values are shown in Table I.
The difference between $\epsilon_\infty$ and $\epsilon'_\infty$
is very small ($\Delta\epsilon_\infty < 10^{-6}t$),\refto{footnote2}
which ensures the reliability of the estimate.
For $n=3/4$, we estimate similarly the ground-state energy by
using 8 and 16 sites clusters with 6 and 12 electrons, respectively.
The obtained $\epsilon_\infty$ is shown in Table II.  
Because in this case we fit the data with two parameters
$\epsilon_\infty$ and $C_1$,
the error is larger than that in the quarter-filled case.
We will use $\epsilon_\infty$ later to compare with
the variational energies.

\fontB\noindent
\beginsection \S3. Comparison with the GWF Near the Supersymmetric Case
\par
\fontA\smallskip
\par\noindent
In this section we compare the results in the previous section
with the Gutzwiller wave function (GWF).  The GWF is defined as
$$
P_d \Phi_{\rm F} = \prod_j (1-n_{j\uparrow}n_{j\downarrow})\Phi_{\rm F},
\eqno{(3.1)}
$$
where $\Phi_{\rm F}$ is a simple Fermi sea.
In 1D case, the analytic expression for the physical quantities were
developed.\refto{Voll1,Voll2}

First let us compare the variational energy.
The expectation value of the kinetic energy is calculated as
$$
E_t = {\langle {\cal H}_t \rangle \over N_{\rm a}}
= -{2t\over 2\pi} \sum_\sigma \int_{-\pi}^\pi dk\cos k
\langle c_{k\sigma}^\dagger c_{k\sigma}\rangle ,
\eqno{(3.2)}
$$
where $\langle \cdots \rangle$ indicates the expectation value in the GWF.
The analytic expression for $\langle c_{k\sigma}^\dagger c_{k\sigma}\rangle$
has been given by an infinite summation.\refto{Voll1}
The detailed calculation is summarized in Appendix A.  Here we show that
the exchange energy can be obtained in a compact form.\refto{Voll2}
 From the expression of $S(k)$ and $N(k)$ (see below and Appendix A), we get
$$
\eqalign{
\langle 4S_i^z S_{i+1}^z \rangle = &- {1 \over \pi}
\bigl\{ {\rm Si}(\pi)-{\rm Si}((1-n)\pi)\bigr\} , \cr
\langle n_i n_{i+1}\rangle  = &n^2 + {1\over 2\pi^2}(\cos 2n\pi -1) \cr
     &+{1 \over \pi}\biggl( {\sin n\pi\over \pi} +(1-n)\cos n\pi\biggr)
\bigl\{ {\rm Si}(\pi)-{\rm Si}((1-n)\pi)\bigr\},
}\eqno{(3.3)}
$$
with $n=N/N_{\rm a}$ being the electron density.
Using a fact that the GWF is singlet,\refto{AndersonSH}
we obtain analytically
$$
\eqalignno{
E_J = &{\langle {\cal H}_J \rangle \over N_{\rm a}} \cr
    = &- {J\over 4\pi}\biggl\{
\bigl\{ 3 + {\sin n\pi\over \pi} +(1-n)\cos n\pi \bigr\}
\bigl\{ {\rm Si}(\pi)-{\rm Si}((1-n)\pi)                      \bigr\} \cr
      &+ \pi n^2 + {1\over 2\pi}(\cos 2n\pi -1) \biggr\}.
&(3.4)}
$$

In Fig.\ 3, we compare the total energy of the GWF with that
of the Bethe Ansatz (BA) at $J/t=2$.
It is surprising that two results are indistinguishable for
any value of $n$ in the scale of this figure.
In fact, for the half-filled or the Heisenberg case ($n=1$),
the energy of the GWF,\refto{Gros,Voll1}
$$
E({\rm GWF})/t=-{1 \over 2}[{3 \over \pi}
{\rm Si}(\pi)+1]=-1.384235\cdots,
\eqno{(3.5)}
$$
is extremely close to that of the Bethe-Ansatz result,
$$
E({\rm BA})/t=-2\ln 2=-1.386294\cdots .
\eqno{(3.6)}
$$
For $n=0.5$, analytic expressions (3.2) and (3.4) give
$$
E_t/t=-0.574632\cdots ,
\eqno{(3.7)}
$$
$$
E_J/J=-0.164230\cdots .
\eqno{(3.8)}
$$
Thus the total energy of the GWF becomes
$$
E({\rm GWF})/t=-0.903092\cdots ,
\eqno{(3.9)}
$$
which is quite close to the exact one:
$$
E({\rm BA})/t=-0.903649\cdots.
\eqno{(3.10)}
$$
The difference is only 0.06\%, which is better than the half-filled
case (0.15\%).

In the low density region, we find that the coincidence is much better.
We can see that the ground-state wave function for 2-electron
problem at $J/t=2$ is given by
$$
|\Psi\rangle = P_d c_{k=0\uparrow}^\dagger
c_{k=0\downarrow}^\dagger |0\rangle .
\eqno{(3.11)}
$$
This means that the GWF is exact in the low density limit.
In fact, the total energies of the GWF at $J/t=2$ is
$$
E({\rm GWF})/t = -2n+{\pi^2 \over 12}n^3 + O(n^4),
\eqno{(3.12)}
$$
which coincides with the BA results up to the order
of $n^3$.  The detailed calculation for (3.12) is also given
in Appendix A.
This is consistent with the fact that the critical exponent $K_\rho$
approaches 1 for $n\rightarrow 0$.
This Fermi liquid state is nothing but the GWF.

In Fig.\ 4, we compare $n(k)$ for the GWF with that for the small
cluster calculation.
As for $n(k)$, two results almost coincide quantitatively except
for the singularity at $k_{\rm F}$.
It is remarkable that the ground state of the $t$-$J$ model has an
enhancement of $n(k)$ in the vicinity of $\pi$, which was considered
before as a pathological behavior of the GWF.\refto{YS1}
It can be shown that this enhancement originates from the
correlated electron motion,
$$
\langle c_{i\sigma}^\dagger c_{j\sigma}
(1-n_{i-\sigma})(1-n_{j-\sigma})\rangle_0 ,
\eqno{(3.13)}
$$
where $\langle \cdots \rangle_0$ represents the expectation value in the free
Fermi see $\Phi_{\rm F}$ without the Gutzwiller projection.
After a straightforward calculation, (3.13) becomes
$$
P_{ij\sigma}(1-n_{-\sigma}+n_{-\sigma}^2+\delta_{ij}n_{-\sigma}
             -P_{ij-\sigma}P_{ji-\sigma}),
\eqno{(3.14)}
$$
and its Fourier transform is
$$
n_{\sigma}^0(k)-2n_{-\sigma}n_{\sigma}^0(k)+f_{2}(k).
\eqno{(3.15)}
$$
Here $P_{ij\sigma}$ is the Fourier transform of
$n_{\sigma}^0(k)=\theta(k_{\rm F}-|k|)$, and $f_{2}$ is given in
Appendix A as one of the lowest order terms in the analytic calculation.
The enhancement of $n(k)$ in the region of $k>k_{\rm F}$
is roughly reproduced from this simple lowest-order, (3.13).

The slight difference between the GWF and the exact ground state is
the behavior around $k_{\rm F}$ which
originates from the TL liquid nature.
The GWF is essentially a Fermi liquid state with discontinuity
$q=\sqrt{1-n}$ at $k=k_{\rm F}$.
On the other hand, $n(k)$ in the true ground state has a
power-law singularity as
$$
n(k)=n(k_{\rm F})-c|k-k_{\rm F}|^\alpha{\rm sgn}(k-k_{\rm F})
\eqno(3.16)
$$
with
$
\alpha=(K_\rho + 1/K_\rho - 2)/4.
$
For $n=0.5$, $K_\rho$ is about 0.85 and is not far from
1 of the Fermi liquid case.\refto{Ogata,Assaad}
This is the reason for the two results are relatively close to
each other even in the vicinity of $k_{\rm F}$.
In order to describe these power-law behaviors around $k_{\rm F}$,
it seems
necessary to introduce low energy excited states
around the Fermi surface into the trial wave function,\refto{YS2}
or a long-range Jastrow factor\refto{Hellberg1,Hellberg2}
as discussed in \S5.

In Fig.\ 5, $S(k)$ and $N(k)$ are compared with the GWF.
$S(k)$ increases for $k<2k_{\rm F}$ and becomes constant for $k>2k_{\rm F}$:
$$
S(k)=\cases{
   -\ln\bigl(1-{|k| \over \pi}\bigr)\ , &\qquad $0\leq k<2k_{\rm F}$  \cr
   -\ln(1-n)\ ,                         &\qquad $2k_{\rm F}<k\leq\pi$ \cr
}\eqno(3.17)
$$
($2k_{\rm F}= n\pi$).  $N(k)$ has different curves
$$
N(k)=\cases{
    {1 \over \pi}
   -{|k| \over 2\pi}\ln {\bigl(1-n+{|k| \over \pi}\bigr) \over (1-n)}\ ,\cr
                 \qquad\qquad\qquad\qquad\qquad
                 (0\leq k<\min (2k_{\rm F}, 2\pi-4k_{\rm F}))  \cr
    {|k| \over \pi}
   -{|k| \over 2\pi}\ln {\bigl(1+n-{|k| \over \pi}\bigr) \over (1-n)}
   + \ln \bigl(1+n-{|k| \over \pi}\bigr)\ , \cr
                 \qquad\qquad\qquad\qquad\qquad
                 (2k_{\rm F}< k< \min (4k_{\rm F}, 2\pi-4k_{\rm F}) \cr
                 \qquad\qquad\qquad\qquad\qquad
                 {\rm when}\ 2k_{\rm F}<2\pi-4k_{\rm F}, {\rm or}\ n<2/3 )\cr
      \ \cr
    2-2n
   +{|k| \over 2\pi}\ln {\bigl(n-1+{|k| \over \pi}\bigr) \over
                         \bigl(1-n+{|k| \over \pi}\bigr) }\ , \cr
                 \qquad\qquad\qquad\qquad\qquad
                 (2\pi-4k_{\rm F}<k<2k_{\rm F}                  \cr
                 \qquad\qquad\qquad\qquad\qquad
                 {\rm when}\ 2\pi-4k_{\rm F}<2k_{\rm F}, {\rm or}\ 2/3<n<1) \cr
      \ \cr
    2n + \ln (1-n)\ , \cr
                 \qquad\qquad\qquad\qquad\qquad
                 (4k_{\rm F}<k<\pi                              \cr
                 \qquad\qquad\qquad\qquad\qquad
                 {\rm when}\ 4k_{\rm F}<\pi, {\rm or}\ n<1/2) \cr
      \ \cr
    2-2n
   +{|k| \over 2\pi}\ln {\bigl(n-1+{|k| \over \pi}\bigr) \over
                         \bigl(1+n-{|k| \over \pi}\bigr) }
   + \ln \bigl(1+n-{|k| \over \pi}\bigr)\ . \cr
                 \qquad\qquad\qquad\qquad\qquad
                 (\max (2k_{\rm F}, 2\pi-4k_{\rm F})<k<\pi  \cr
                 \qquad\qquad\qquad\qquad\qquad
                 {\rm when}\ \pi < 4k_{\rm F}, {\rm or}\ 1/2<n<1) \cr
}\eqno(3.18)
$$

Figures 4 and 5 show that the GWF is an extremely good trial
function to describe globally the 1D $t$-$J$ model in the supersymmetric
case, although it is not the exact ground state.
The $J$-dependence of the variational energy for the GWF is also shown
in Fig.\ 2 using (3.7) and (3.8).  It is apparent that
the GWF is very good only in the vicinity of the supersymmetric case.
\par

It is possible to speculate the reason.
An up-spin electron, for example, can hop to a neighboring site
with the same magnitude $t=J/2$ independent of the state of the
neighboring site (hole or down-spin electron).\refto{tjBA2}
This fact makes the up-spin part of
the wave function similar to the Fermi sea where the
up-spins move freely without any correlation to the down-spins.
Actually the electron system without any interaction has an
energy
$$
E_t/t = -{4 \over \pi}\sin{{\pi n}\over 2}
=-2n+{{\pi^2} \over {12}}n^3+O(n^5),
$$
which coincides with the exact result up to $O(n^3)$.
\par

Now we discuss the flat region of $S(k)+N(k)$ in
$k>2k_{\rm F}$ which we have shown in the previous chapter (Fig.\ 1).
The summation of them is rewritten as
$$
\eqalignno{
&\ {1 \over N_{\rm a}}\sum_{j,\ell} \bigl\{
  2\langle n_{j\uparrow}n_{\ell\uparrow}\rangle
+ 2\langle n_{j\downarrow}n_{\ell\downarrow}\rangle
- \langle n_j\rangle \langle n_\ell \rangle \bigr\}
  e^{ik(r_j-r_\ell)} \cr
&= {4 \over N_{\rm a}}\sum_{j,\ell} \bigl\{
  \langle n_{j\uparrow}n_{\ell\uparrow}\rangle
- \langle n_{j\uparrow}\rangle \langle n_{\ell\uparrow}\rangle \bigr\}
e^{ik(r_j-r_\ell)}.
&(3.19)}
$$
Therefore it represents the density correlation between the same
species of spin.
The coincidence of this quantity to the non-interacting value
shows that the each spin behaves freely.
The Gutzwiller projection only affects the correlation
between the different species of spins, such as
$\langle n_{j\uparrow}n_{\ell\downarrow}\rangle $.
The fact that the GWF behaves almost as a free state will
be confirmed also for such quantities as charge and spin
susceptibilities and magnetization curve in \S6.

It seems that the fact the GWF is excellent in the supersymmetric
case does not occur by chance.
Recall that for the 1D supersymmetric $t$-$J$ model with $1/r^2$ long-range
coupling, the exact ground state is the GWF.\refto{Kura,Kato}
The lowest eigenvalue specified by electron density $n$ and
magnetic polarization $m=(N_\uparrow-N_\downarrow)/N_{\rm a}$
consists of independent contributions of $n$ and $m$;
the logarithmic correction does not exist for
the critical behaviors.
It is surprising that except for the long-range behaviors the main
features of this long-range model are shared with the ordinary $t$-$J$
model.
Furthermore recently it has been found that many low-lying
excited states of the above long-range $t$-$J$ model, including the
case of finite magnetic polarization, are represented also by
Gutzwiller-projected determinantal wave functions.\refto{YK,YKO,Ha}
Thus to elucidate the relation between these models and to study
excited states by the variational approach are fascinating
future problems.

\fontB\noindent
\beginsection \S4. Fermi-Liquid-Type Jastrow Functions
\par
\fontA\smallskip
\par\noindent
In this section we study the Fermi-liquid-type wave functions
to describe the $t$-$J$ model away from $J/t=2$, keeping the
application to higher dimensions in mind.
In many-body problems Jastrow-type wave functions with two-body
correlation factors are often used.
First, notice that the trial state becomes non-singlet if the Jastrow
correlation is spin-dependent.\refto{footnote4}
Therefore we study here only
the spin-independent charge density correlation:\refto{footnote3}
$$
\Psi=\prod_{j\ell}\prod_{\sigma\sigma'}
     \Bigl[1-\bigl(1-\eta(r_{j\ell})\bigr)
     n_{j\sigma}n_{\ell\sigma'}\Bigr]\Phi_{\rm F} , \eqno(4.1)
$$
where $r_{j\ell}=|r_j-r_\ell|$.
For $\eta(r)$ simple forms are desirable, satisfying the
condition of the $t$-$J$ model: $\eta(0)=0$.
This condition is to project out the double occupancies.
In this section we consider two cases in addition to the GWF:
$$
\eta(r)=\cases{{2 \over \pi}\arctan{r \over \zeta}
             \qquad\qquad &({\rm a}) \qquad ({\rm RJWF})\cr
                  1
             \qquad\qquad &({\rm b}) \qquad ({\rm GWF})\cr
                  1+{\alpha \over {r^\beta}}
             \qquad\qquad &({\rm c}) \qquad ({\rm AJWF})\cr
                 } , \eqno(4.2)
$$
where $r\neq0$ and $\zeta$, $\alpha$ and $\beta$ are positive
variational parameters.
The typical forms of $\eta(r)$ are shown in Fig.\ 6.
In these function, $\eta(\infty)=1$ holds and the value of
$\eta(1)/\eta(\infty)$ is finite for every parameter set.
This point is essentially different from the function with a
long-range Jastrow factor discussed in the next section.
\par

The form (a), which we call the repulsive Jastrow wave function
(RJWF) in this paper, includes repulsive correlation.
It prefers configurations with electrons mutually apart.
In the limit $\zeta\rightarrow 0$, it is reduced to the GWF (b).
This wave function was previously introduced to study the repulsive
Hubbard model in strong correlation.\refto{YS2}
In that work it was found that the RJWF lowers the variational energy
in the Hubbard model and it reproduces qualitatively the enhancement
at $2k_{\rm F}$ of $S(k)$.
Therefore we expect that the RJWF is suitable also for the small-$J$
region of the $t$-$J$ model.

An attractive Jastrow wave function (AJWF) with the correlation factor
(c) favors local configurations with electrons close to each other.
The parameter $\alpha$ adjusts the amplitude of such attractiveness;
as $\alpha$ increases, more emphasis are laid upon attractive
electron configurations.
On the other hand, the other parameter $\beta$ controls the decaying
behavior of $\eta(r)$. With increasing $\beta$, effective attractive
range becomes narrower.
It is reduced to the GWF when $\alpha\rightarrow0$ or $\beta\rightarrow0$.
We will show that the AJWF properly describes a homogeneous phase for
relatively small $\alpha$ as well as a fully phase-separated state
for large values of $\alpha$.

Comparing with the RJWF and the AJWF, we may regard the GWF as a
``free-electron" state in that there is no amplitude modification from the
non-interacting state except for the exclusion of doubly occupied sites.

We evaluate the expectation values by the variational Monte Carlo
(VMC) method.\refto{Gros,YS1}
This method gives virtually {\it exact} expectation
values for given trial functions.
In the VMC calculations in this work we use systems with electron
number $N=4I+2$ ($I$: integer) with the periodic boundary condition.
Sample numbers ($3\times 10^4\sim 2\times10^5$) and sampling intervals
(50 Monte Carlo steps at the maximum) are taken so as to reduce
statistical
fluctuations enough.
In this section we use typically 60- and 72-site systems.
\par
Before going to the cases of $J/t \neq 2$, we check the supersymmetric case.
Total energy per site of the RJWF and the AJWF for $J/t=2$ and $n=0.5$
are shown in Figs.\ 7(a) and 7(b), respectively.
These figures tell us that the Jastrow correlation factors raise
the variational energy and that the GWF is the best trial state for
$J/t=2$.
In the following we will see that the RJWF has lower energy than the
GWF for $J/t<2$ (\S4.1) and the AJWF has lower energy than the GWF
for $J/t>2$ (\S4.2).
\par\medskip
\noindent
{\it 4.1\quad Repulsive Case}
\par\smallskip\noindent
We starts with a brief survey of energy.
In Fig.\ 8, $E_t/t$ and $E_J/J$ in the RJWF are plotted as a function of
$\zeta$ for $n=0.5$.
As $\zeta$ increases, electrons keep apart from each other and
thus $E_t$ tends to decreases and $E_J$ increases.
We show the total energy $E(\zeta)$ for $J/t=1.0$, as an example,
in Fig.\ 9 using the values in Fig.\ 8.
The minimum is situated at $\zeta\sim 0.7$; the energy is considerably
improved upon the GWF ($\zeta=0$).  Similarly we find that
the minimum appears at $\zeta\sim 1.6$ (for $J/t=0$), 1.2 (0.5),
0.7 (1.0), 0.4 (1.5), and 0 (2.0), respectively.
The repulsive interaction represented by the magnitude of $\zeta$
becomes weaker with increasing $J$.
Variational energies thus obtained are plotted also in Fig.\ 2 to
compare with the exact ground-state energy.

Next, let us consider the critical value $J_{\rm c}$ below
which the RJWF has a lower variational energy than the GWF.
In order to estimate $J_{\rm c}$ we only have to see the
$\zeta$-dependence of the variational energy near $\zeta=0$:
if the slope of the energy at $\zeta=0$ is positive, the optimal
state is $\zeta=0$, which is the GWF.
On the other hand, if the slope at $\zeta=0$ is negative,
the variational energy has a minimum at a certain value of $\zeta>0$
and the RJWF is more stable than the GWF ($\zeta=0$).

In Fig.\ 10 variational expectation values, $E_t/t$ and $E_J/J$,
in the RJWF are shown for $n=0.3$ and small $\zeta$.
 From this figure we read
$E_t(\zeta)/t= -0.471  - 0.0413\zeta$ and
$E_J(\zeta)/J= -0.0540 + 0.0217\zeta$.
Since the total variational energy is given by
$$
E(\zeta)=E_t(\zeta) + E_J(\zeta), \eqno(4.3)
$$
the slope of $E(\zeta)$ near $\zeta=0$ is
$-0.0413 t + 0.0217 J$.
This means that for $J/t< 0.0413/0.0217\simeq 1.9$,
the slope is negative and the RJWF improves the variational energy
upon the GWF.
In a similar way, we obtain the value of $J_{\rm c}/t$ also for
$n=0.5$, 0.75 and $0.833\cdots$.
As $n$ increases, the statistical fluctuation becomes severe
and it is not easy to estimate accurate values.
However, we find $J_{\rm c}/t=2.0\pm 0.2$ for the above densities.
 From this we conclude that a repulsive correlation is
necessary to reproduce the ground state for $J/t<2$.
Actually the value of $J_{\rm c}/t$ ought to be slightly shifted
to a larger value of $J/t$ in the high density region, as will be
seen in the next section.
This is consistent with the diagonalization study.\refto{Ogata}
\par

Leaving the discussion of correlation functions in the next section,
here we focus on the discontinuity
$q$ of momentum distribution $n(k)$ at $k_{\rm F}$, which is
one of the characteristic properties of a Fermi-liquid-type
wave function.
This quantity is equal to the wave function renomalization
factor and related to the inverse of the
effective mass $m^*/m$ in the Fermi liquid theory.
Similarly it is the energy reduction factor in the Gutzwiller
approximation.
In Fig.\ 11 the value of $q$ is plotted as a function of $n$
calculated with the optimized wave function for each value of $J/t$.

For the supersymmetric case the result of the GWF,\refto{Voll1}
$q=\sqrt{1-n}$, is shown.
In this case $q$ becomes 1 in the limit of $n\rightarrow 0$,
which is the value of the non-interacting system.
In the other limit $n\rightarrow 1$, $q$ vanishes.
This implies the Fermi surface vanishes, namely
effective mass diverges, corresponding to the metal-insulator
transition.
As $J/t$ decreases, $q$ becomes small; the correlation
affects the whole electron density in this case.
Although the result for $J/t=2.5$ includes relatively large
statistical error, it is obvious that $q$ tends to decrease,
with increasing $J/t$ from $J/t=2.0$.
Thus we conclude that the supersymmetric case is the most
weakly interacting; as $J/t$ goes away from it, the correlation
effect becomes severer.
\par\medskip\noindent
{\it 4.2 Attractive Case}
\par\smallskip\noindent
We study the AJWF in this subsection.
For $J/t>2$, it is natural to expect that the attractive correlation
between electrons is dominant.
In Fig.\ 12 we show the variational expectation values of ${\cal H}_t$
and ${\cal H}_J$ for the AJWF ($n=0.75$).
$E_J$ becomes lower and $E_t$ becomes higher as $\alpha$ increases,
because the amplitude of configurations with electrons located next
to one another increases.
Note that $E_t$ abruptly approaches to zero and $E_J$ to the energy
of the spin system in the region $\alpha>10$ for $\beta=0.625\sim1$.
This is nothing but a sign of the phase transition.
We will come back to this point shortly.

First we estimate the critical values above which the GWF becomes
unstable against the AJWF by studying the small-$\alpha$ (and various
$\beta$) behavior, as in the RJWF case.
We obtain $J_{\rm c}/t=2.0\pm 0.1$ for $n=0.5$ and $2.0\pm 0.4$
for $n=0.25$ and 0.75.
\par

Searching the energy minimum in the $\alpha$-$\beta$ plane, we
obtain the optimized energy for $J/t>2$.
The cases of $n=0.5$ are included in Fig.\ 2.
For $n=0.75$ and $J/t=3.3$, 3.4 and 3.5, the total
energies as a function of variational parameters are shown in
Figs.\ 13(a)-(c), respectively.
The energy is improved on that of the GWF in every case.
The value of $\alpha$ giving the energy minimum becomes large
with increase of $J/t$; this means the enhancement of the
attractive correlation.

One can read that the aspects of the minima are quite different
among three figures of Figs.\ 13(a)-(c).
For simplicity, let us consider the case of $\beta=1$ (empty circle).
In each figure the energy curve of $\beta=1$ has two local minima,
that is, one around $\alpha=1\sim 2$ (minimum H) and one around
$\alpha=15$ (minimum S).
In Fig.\ 13(a) the lowest energy is given by the minimum H, and
the minimum S has a higher value.
In Fig.\ 13(b) the two minima have comparable values.
On the other hand, for $J/t=3.5$ (Fig.\ 13(c)) the situation is
opposite to Fig.\ 13(a).
Thus we find a switching from the state of the minimum H to one of the
minimum S around $J/t=3.4$.

Let us see that the optimal state for $J/t>3.4$ represents the
phase separation.
First, the snapshots of electron configurations taken along a
Monte Carlo (MC) sweep (Fig.\ 14) show a formation of an
electron cluster.
Although small density fluctuation can be seen near the boundary,
this edge effect will be irrelevant as $N_{\rm a}\rightarrow\infty$.
Second, the minimized energy for $J/t>3.4$ is quite close to the
energy of the Heisenberg chain corrected by electron density (Fig.\ 2).

To see this more quantitatively, charge-density correlation
function in real space,
$$
N_r={1 \over N_{\rm a}}\sum_i[\langle n_in_{i+r}\rangle - n^2],
\eqno(4.4)
$$
is plotted in Fig.\ 15 for $\beta=0.75$.
For $\alpha=0$ (GWF),
$N_r$ has a negative dip in a short distance due to the exchange
hole of the original Fermi sea, and falls off as $r^{-2}$ in a
long distance and converges to zero.
When $\alpha$ is switched on, the effect of the exchange hole is
compensated by the weak attractive correlation; on the other hand,
long range part of $N_r$ hardly changes.
As the value of $\alpha$ increases further, $N_r$ abruptly changes its
behavior at $\alpha=10$.
It approaches the value of the completely separated phase
($\alpha=\infty$) expressed as
$$
N_r=\cases{n-n^2-r/N_{\rm a}&$(0\leq r/N_{\rm a}\le
              1-n)$\cr
           2n-n^2-1&$(1-n\leq
              r/N_{\rm a}\le 1/2)$\cr
    }\qquad {\rm for}\ {1 \over 2}\leq n\le 1\ \eqno(4.5a)
$$
$$
N_r=\cases{n-n^2-r/N_{\rm a}&$(0\leq r/N_{\rm a}\le
              n)$\cr
           -n^2&$(n\leq r/N_{\rm a}\le 1/2)$\cr
    }\qquad {\rm for}\ 0\leq n\leq {1 \over 2}.\ \eqno(4.5b)
$$
In Fig.\ 16 $N_1$, $N_{r=N{\rm a}/2}$ and
$\delta_{N{\rm a}/4}\equiv -N_{\rm a}\bigl(dN_r/dr\bigr)$ for
$r=N_{\rm a}/4$ are plotted versus $\alpha$.
The transition at $\alpha\sim 10$ can be seen.
When a large cluster of electrons is formed, $N_1$ and $N_{r=N{\rm a}/2}$
are most influenced by the density fluctuation near the boundary.
Inversely $\delta_r$ for $r\sim N_{\rm a}/4$ depends mainly on the
fluctuation in the middle of the electron cluster, so that it
represents the phase transition sharply.

In Fig.\ 17 $S(k)$ is shown for some values of $\alpha$.
For $\alpha=0$ it has a cusp at $k=2k_{\rm F}= \pi/2$.
As $\alpha$ becomes large $S(\pi)$ is enhanced, showing
antiferromagnetic correlation for the Heisenberg chain.
For $\alpha=30$, $S(k)$ of the AJWF is considerably similar to
the Heisenberg spin system corrected with $n$.
The behavior of $S(k)$ is also consistent with the phase
separation.

In this way we find the phase boundary to the phase separation
is at $J/t=3.4$ for $n=0.75$.
Similarly we obtain $J/t\simeq3.2$ ($n=0.25$) and $J/t\simeq3.3$ ($n=0.5$).
These values are in good agreement with the exact-diagonalization
result.\refto{Ogata}
\fontB\noindent
\beginsection \S5. Tomonaga-Luttinger-Liquid-Type Jastrow Function
\par
\fontA\smallskip
\par\noindent
In this section we study the trial state proposed by Hellberg
and Mele.\refto{Hellberg2}
For the variational states treated in the previous section,
the correlation exponent is always the same as the Fermi liquid
($K_\rho=1$).
This is because the Jastrow factor $\eta(r)$ used in (4.2) approaches
1 rapidly, namely is short ranged.
If the correlation factor $\eta(r)$
is long-ranged, the correlation exponent becomes non-trivial which
is consistent with the TL liquid behavior.
Hellberg and Mele\refto{Hellberg2} showed that a variational state
$$
|F(r_{i\uparrow}, r_{j\downarrow})|^\nu \Phi_{\rm F},
\eqno{(5.1)}
$$
gives non-trivial exponents,
where $|F(r_{i\uparrow}, r_{j\downarrow})|$ is a Slater determinant
of all the electron positions.
Actually the correlation exponent is related to the variational
parameter $\nu$ as\refto{KH}
$$
K_\rho={1\over 2\nu + 1},
\eqno{(5.2)}
$$
which was derived using the asymptotic Bethe Ansatz and
the scaling relations of the TL-liquid theory.

If we rewrite this wave function
in the form (4.1) using the Vandermonde's determinant identity,
the corresponding Jastrow correlation can be written as
$$
\eta(r)=\biggl[ {N_{\rm a}\over\pi}\sin \bigl({\pi\over N_{\rm
a}}r\bigr)\biggr]^\nu .
\eqno{(5.3)}
$$
When $\nu>0$, this correlation is repulsive and is attractive for $\nu<0$.
As we can see from (5.3), the factor $\eta(r)$ behaves as $N_{\rm a}^\nu$
at the longest distance, $r=N_{\rm a}/2$.  It depends on the
system size $N_{\rm a}$, and thus it is a very long-ranged Jastrow factor which
is different from the conventional Jastrow factor, (4.2).
This long-range behavior stands for the effect of phase shift
of two-particle collisions and eventually changes the
correlation exponent from the non-interacting value.

In Fig.\ 18, we show the phase diagram obtained using this trial
state.  The solid lines represent the contours of constant
correlation exponent $K_\rho$.
Comparing this phase diagram with that obtained from the
exact diagonalization of small clusters,\refto{Ogata}
we find that the variational wave function is very good in
the vicinity of $J/t=2$ including the correlation exponents.
However agreement is not so good near $J/t\sim 0$.
For example, the optimized variational parameter $\nu$ for $J/t=0$ is
$\nu=1, (\nu=0.75, \nu=0.5)$ near $n=0, (n={1\over 2}, n=1)$,
respectively, while the exact exponent is
$K_\rho={1\over 2}$, which corresponds
to $\nu={1\over 2}$, regardless of $n$.
One can see this deviation directly by examining the correlation
functions.
Since the density correlation function $N(k)$ is exactly the same
as that of the free spinless fermions at $J/t=0$, there must not
be a peak at $2k_{\rm F}$.
However $N(k)$ calculated with $\eta$ of (5.1) has a small peak
at $2k_{\rm F}$ owing to the deviation of the exponent (Fig.\ 19(a)).

To understand the reason of this deviation, we rewrite the
variational state (5.1) as\refto{KuramotoPC}
$$
\eqalignno{
|F(r_{i\uparrow}, r_{j\downarrow})|^\nu \Phi_{\rm F},
 &=\prod_{i<j} |z_i-z_j|^\nu \prod_{i<j} |z_{i\uparrow}-z_{j\uparrow}|
   \prod_{i<j} |z_{i\downarrow}-z_{j\downarrow}| \cr
 &=\prod_{i<j} |z_i-z_j|^{1+\nu} / \prod_{i,j} |z_{i\uparrow}-z_{j\downarrow}|,
&{(5.4)}
}
$$
apart from the sign.  Here $z_{j\sigma}=\exp(2\pi ir_{j\sigma}/N_{\rm a})$
and $\{ z_j\} $ represents all the electrons irrespective of their spins.
On the other hand, the exact density correlation function at $J/t=0$ is
reproduced by the free spinless fermion, $\prod_{i<j} |z_i-z_j|$.
We can see that it is difficult to reproduce
the exact density correlation using the form (5.4).

\refis{KuramotoPC} Suggested by Y.\ Kuramoto.

In this connection, for the region of $J/t\sim 0$, a better trial
function will be obtained as the form, $\Psi=\chi\Phi_{\rm SF}$
on the analogy of the exact eigenfunction in the small-$J/t$ limit,
\refto{OgataShiba} where $\Phi_{\rm SF}$ is the wave function of
spinless fermion and $\chi$ is that for spin degree of
freedom.\refto{OgataKoba}
\par

Next we make a comparison of correlation functions.
Figure 19(a) shows $n(k)$, $S(k)$ and $N(k)$ for the optimized
wave function (5.1) at $J/t=0$ and $n=0.5$,
together with the result of exact diagonalization.
The reproducibility of the global features is not so good.
In the same figure we compare the correlation functions calculated
in the optimized RJWF; the two variational states give almost
identical results.
Actually we find the two optimized forms of $\eta(r)$ of
two variational states are similar in short distance.
This suggests that the global features of the
correlation functions are determined mainly by the short-range
behavior of the Jastrow factor.

Figures 19(b) and 19(c) show similar comparisons between the
variational states and the exact diagonalization at $J/t=1$ and
2.6, respectively.
Since the values of $J/t$ is nearer to 2, three results are
close to each other except for the vicinity of $k_{\rm F}$ in $n(k)$.
\par

Before closing this section we confirm that the long-range part
of the correlation
factor controls the critical behavior of correlation functions,
while the short-range part determines the global features of
correlation functions as well as energy.
Let us consider the wave functions with four different types
of correlation factors, two of which connect the correlation
factors of two kind of variational states.
Namely,
\par\smallskip\qquad\qquad\qquad case 1)\qquad
$\eta(r)=\eta_{\rm TLL}(r)$\qquad\qquad for all $r$,
\par\smallskip\qquad\qquad\qquad case 2)\qquad
$\eta(r)=\cases{\eta_{\rm FL}(r), &\qquad for $r<r_{\rm c}$; \cr
                \eta_{\rm TLL}(r), &\qquad for $r\ge r_{\rm c}$, \cr}$
\par\smallskip\qquad\qquad\qquad case 3)\qquad
$\eta(r)=\cases{\eta_{\rm TLL}(r), &\qquad for $r<r_{\rm c}$; \cr
                \eta_{\rm FL}(r), &\qquad for $r\ge r_{\rm c}$, \cr}$
\par\smallskip\qquad\qquad\qquad case 4)\qquad
$\eta(r)=\eta_{\rm FL}(r)$\qquad\qquad\quad for all $r$,
\par\smallskip\noindent
where $\eta_{\rm TLL}$ denotes $\eta(r)$ of (5.3),
$\eta_{\rm FL}$ either that of the RJWF or of the AJWF (4.2), and
$r_{\rm c}$ is a certain value of $r$ which divides $\eta(r)$
into a short-range part and a long-range part.
We choose the variational parameters and $r_{\rm c}$ so that
the two parts may be smoothly connected.
By using these correlation factors, we perform VMC calculations for
the energy, $n(k)$, $S(k)$ and $N(k)$.
We fix the system size to $N_{\rm a}=100$ for $n=0.5$ and take
$5\times10^4$ samples for each case.
\par

For the repulsive case we choose $r_{\rm c}=5$, $\nu=0.5$
(the optimal value for $J/t\sim 0.7$) and $\zeta=6.9847$.
In Table III(a) we show expectation values of several
physical quantities for the four cases.
$\delta n(k)$ in the last two columns indicates the differences of
$n(k)$ between the two possible $k$-points adjacent to $k_{\rm F}$
from inside and outside, respectively.
For the bulk quantities like $E_t/t$, $E_J/J$ and $n(k=0)$, the cases
1) and 3) give similar values, which are different from the cases
2) and 4).
This implies that the short-range part of the correlation factor
is responsible for the bulk properties like energy and the magnitude
of correlation functions.
On the other hand, for $\delta n(k)$ the cases 1) and 2) have
similar values compared with the cases 3) and 4).
This indicates the long-range part controls
the critical behavior of correlation functions.
\par
In Table III(b), the result of the same quantities in the
attractive regime is summarized.
Here we select $r_{\rm c}=10$ and $\nu=-0.15$ (the optimal value
for $J/t\sim 2.5$), and the two parameters are determined by requesting
$\eta_{\rm TLL}(1)=\eta_{\rm AJ}(1)$ and
$\eta_{\rm TLL}(r_{\rm c})=\eta_{\rm AJ}(r_{\rm c})$
as $\alpha=0.6804$ and $\beta=0.5481$.
The tendency agrees with the repulsive case.
These facts explain the similarity of $n(k)$ and the slight
energy difference shown in Figs.19.
\par
\fontB\noindent
\beginsection \S6. Static Properties of Charge and Spin
\par
\fontA\smallskip
\par\noindent
In this section we focus on the behavior of charge susceptibility
$\chi_{\rm c}$ and spin susceptibility $\chi_{\rm s}$.
Since the trial states in the half filling are always Mott-insulating
for the $t$-$J$ model,
we can see the critical behavior variationally,
and discuss the difference between the present case and the
Brinkman-Rice transition (Appendix B).
\par
We calculate $\chi_{\rm c}$ and $\chi_{\rm s}$ from the VMC data
for systems with finite sizes from the formulae:
$$
\chi_{\rm c}^{-1}={{\partial^2 E}\over{\partial n^2}}
          ={{N_{\rm a}^2}\over{4}}\{E(N+2,0)+E(N-2,0)-2E(N,0)\}, \eqno(6.1)
$$
and
$$
\chi_{\rm s}^{-1}={{\partial^2 E}\over{\partial m^2}}
          ={{N_{\rm a}^2}\over{16}}\{E(N,8)+E(N,0)-2E(N,4)\}, \eqno(6.2)
$$
where $E(N,M)$ is total energy per site of $N$-electron systems
with $M=N_{\uparrow}-N_{\downarrow}$.
Henceforth we take $1/t$ as the unit of $\chi_{\rm c}$ and
$\chi_{\rm s}$.
We confirm that the size dependence is mostly negligible
for the systems we use: $N_{\rm a}=100\sim 200$.
In contrast to the Hubbard model in the strong coupling regime,
relatively accurate estimate of $\chi$ is possible for the
1D $t$-$J$ model because of the less statistical fluctuations.
In this section, we mainly use the wave function (5.3) and GWF.
Typically $5\times10^4\sim 2\times10^5$ independent VMC samples
are used for each value of the parameter $\nu$.
To search the optimized value of $E(N,M)$ for each $N$ and $M$,
we pick out 61$\sim$121 values of $\nu$ between $-1$ and 2.
\par\medskip

\noindent{\it 6.1\quad Charge Susceptibility}
\par\smallskip\noindent
In Fig.\ 20 $\chi_{\rm c}$ is shown for several values of $J/t$.
Here we observe that with increasing $J/t$, $\chi_{\rm c}$ becomes
large.
This is because enhanced attractive interaction between electrons
enlarges charge compressibility $\kappa$ ($=\chi_{\rm c}/n^2$).
In the high electron density region ($n\sim 1$) $\chi_{\rm c}$ for every
value of $J/t$ is divergent as $n\rightarrow 1$.
This divergence is due to the strong correlation effect, which
suppresses charge fluctuation;
this is in contrast with the non-interacting case,
where $\chi_{\rm c}^{-1}=\pi\sin(n\pi/2)$ and remains a finite value
$1/\pi$ as $n$ approaches 1.
For $J/t=0$ and 2 the exact values are known; the spinless fermion
result $\chi_{\rm c}^{-1}=2\pi\sin(n\pi)$ for the former case, and
the Bethe Ansatz solution for the latter.\refto{tjBA2,KY2}
In both cases the exact $\chi_{\rm c}$
diverges as $\chi_{\rm c}\propto 1/\delta$ ($\delta=1-n$) in
the limit of $n\rightarrow 1$.
The variational result is quantitatively similar to the exact one.
To see this divergence more closely, we plot chemical potential
$\mu=\partial E/\partial n$ vs. $\delta^2$ for $J/t=0$ and
2 in Fig.\ 21.
For $J/t\le 2$ we can fit $\mu$ as $\mu =\mu_0-a\delta^2$ in the
vicinity of the half filling, although the linearity is not so clear
for $J/t>2$.
Since $\chi_{\rm c}=\partial n/\partial \mu$,
the coefficient $a$ is related to the divergence of $\chi_{\rm c}$
as $\chi_{\rm c}=2a/\delta (\delta\rightarrow 0)$.
 From Fig.21, we determine this coefficient $2a$ for a couple
of values of $J/t$, which are shown in Table IV.
\par

As can be seen in Table IV, the estimation by the variational
state is in good agreement with the exact value at $J/t=0$.
However at $J/t=2$, the deviation is quite large.
This is probably because the range of linear behavior of $\mu$
vs. $\delta^2$ is rapidly reduced as $J/t$ increases.\refto{KY2}
\par

The divergence of $\chi_{\rm c}$ near the Mott transition has been
understood from the Bethe Ansatz.
Using the correlation exponent $K_\rho$ and the charge velocity
$v_{\rm c}$, $\chi_{\rm c}$ can be written
as\reftorange{Haldane}{Schulz,FK}{KY1}
$$
\chi_{\rm c}={{2}\over{\pi}}{{K_\rho}\over{v_{\rm c}}}.
$$
Since $v_{\rm c}$ vanishes linearly as $\delta\rightarrow 0$,
$v_{\rm c}=b\delta$, $\chi_{\rm c}$ diverges as\refto{Kawakami3,FI,Imada}
$$
{{2}\over{\pi}}{{K_\rho}\over{b}}{{1}\over{\delta}}.
$$
Since $K_\rho\rightarrow 1/2$ as $\delta\rightarrow 0$\refto{Ogata}
the coefficient $2a$ is solely related to the coefficient $b$, which is
$2\pi\ge b\ge 3\pi\zeta(3)/(16(\ln 2)^2)$
for $0\le J/t\le 2$.\refto{Ogata,KY2}
However, for $J/t>2$, we cannot expand $\mu$ as $\mu=\mu_0-a\delta^2$,
but fit $\mu$ as $\mu=\mu_0-a\delta^p$ instead.
The leading power $p$ is estimated as $1<p<2$, namely $p=1.8$
(for $J/t=2.0$), 1.5 (2.5) and 1.3 (3.0).
Although $\chi_{\rm c}$ diverges as $\delta\rightarrow 0$ for
$J/t\ge 2$, we have not understood the origin of this power.
This may be due to the poorness of the trial wave function of
(5.1) or due to an anomalous dependence of $v_{\rm c}$ as
$v_{\rm c}\sim a\delta^{p-1}$.
At any rate, we can consider that this divergence is attributed
to the divergence of density of state at the band edge of the
spinless fermion.
\par

In low electron density area ($n\sim 0$) $\chi_{\rm c}$ again diverges
except for the case of $J/t=2.5$, in which the variational state
becomes unstable against phase separation and $\chi_{\rm c}$
becomes negative.
One can find that the divergence of $\chi_{\rm c}$ is proportional to
$1/n$ by plotting $\mu$ as a function of $n^2$.
The coefficients of $1/n$ are shown also in Table IV.
Since for $J/t=2$ the trial wave function (GWF) is exact in the
limit $n\rightarrow 0$, the exact value is within the range of
error.
Furthermore, the GWF result and the exact one are extremely close to
the non-interacting gas, which means a ``free electron" state
is realized in the supersymmetric case.
Actually using the variational energy (3.12), we obtain
$$
\chi_{\rm c}={{2}\over{\pi^2n}}+O(n^0),
$$
which is the same as in the non-interacting case.
As $n$ increases, however, the suppression of charge fluctuation in
the $t$-$J$ model prevents the wave function from the free-electron
behavior.

For $J/t=0$ the variational value is a little
different from the exact one.
This difference corresponds to the deviation of $K_\rho$ in this
area, as described in the previous section.
The divergence of $\chi_{\rm c}$ is not due to the correlation effect,
but simply to the divergence of the density of state at the band
edge.
\par\medskip
\noindent{\it 6.2\quad Magnetic Properties}
\par\smallskip\noindent
First, we discuss spin susceptibility under zero field.
Figure 22 shows $\chi_{\rm s}$ for a couple of values of $J/t$.
In sharp contrast to $\chi_{\rm c}$, $\chi_{\rm s}$ does not
diverge as $n\rightarrow 1$ for every value of $J/t$ but remains
a finite value close to that of the Heisenberg chain.\refto{Griffiths}
The divergence of $\chi_{\rm c}$ hardly affects $\chi_{\rm s}$;
this is due to the separation between spin
and charge degrees of freedom in the low-energy excitations.
\par
For $J/t=0$, all the spin configurations are degenerate, hence
$\chi_{\rm s}$ is infinite.
This aspect is special in 1D.
By introducing $J/t$, this degeneracy is lifted and
the value of $\chi_{\rm s}$ becomes finite except for the
low density limit, which is shown in Fig.\ 22.
In the TL liquid theory, $\chi_{\rm s}$ can be represented
as\refto{Haldane,Schulz,Kawakami3,FK}
$$
\chi_{\rm s}={{2}\over{\pi}}{1\over{v_{\rm s}}}
$$
where $v_{\rm s}$ is the spin velocity.
Basically $v_{\rm s}$ is proportional to the exchange coupling $J$
and thus $\chi_{\rm s}$ decreases as $J/t$ increases, which means
that the enhanced exchange coupling hinders the response of spins
to the magnetic field.
For a low density area ($n\sim 0$) $\chi_{\rm s}$ diverges since
$v_{\rm s}\propto n$, which is due to the divergence of density
of states.
On the other hand as $n\rightarrow 1$, $\chi_{\rm s}$ approaches
that of Heisenberg chain,
$$
\chi_{\rm s}\rightarrow {{2}\over{\pi}}{{1}\over{v_{\rm sw}}}
$$
with $v_{\rm sw}$ being the spin wave velocity of Heisenberg
chain: $v_{\rm sw}=J\pi/2$.
\par
For the supersymmetric case, we observe that the variational results
agree quite well with that of the non-interacting system
(Pauli paramagnetism) $\chi_{\rm s}^{-1}=\pi\sin(n\pi/2)$,
especially in low density region.
Here the idea of the GWF as a \lq\lq free electron" state is again useful.
Like $\chi_{\rm c}$, however, the two results become a little
apart as $n\rightarrow 1$.
For $J/t=2.5$ and $n<0.1$, the decrease of $\chi_{\rm s}$ is
due to the phase separation.
\par
Now let us turn to the case under a finite magnetic
field $g\mu_{\rm B}H$, and here we put $g=2$ and $\mu_{\rm B}=1$.
Applying an external field along $z$-axis, a Zeeman term,
$$
{\cal H}_{\rm ext} = -2H\sum_j S^z_j = -N_{\rm a}Hm, \eqno(6.3)
$$
is added to the original Hamiltonian (2.1).
Total energy per site is written as,
$$
E=E(m,n)-Hm, \eqno(6.4)
$$
where $E(m,n)$ is total energy per site with electron density
$n$ and magnetization $m$ under zero field.
\par
In Fig.\ 23 $E(m,n)$ is shown for $n=0.5$ and 1.0 as a function
of $m$ ($0\le m\le n$).
For each case the energy is a monotonically increasing
function of $m$.
By minimizing the total energy (6.4), we obtain the magnetization
curves, which are shown in Fig.\ 24.
The critical field $H_{\rm s}$ at which spin saturates is
determined from the slope of $E(m,n)$ near $m=n$.
The value of $H_{\rm s}$ is 0.34($J/t=1.0$), 0.99(2.0) and 1.32(2.5)
for the quarter filling, and 2.000(2.0) for the half filling.
\par
For the half filling, the data of the GWF is very close to the
exact value (Heisenberg antiferromagnet)\refto{Griffiths} for all
the range of $H$.\refto{footnote5}
In the weak field limit, $m/H$ is nothing but $\chi_{\rm s}$,
the similarity of the two results is obvious from Fig.\ 22.

For the quarter filling, $m$ saturates at smaller $H$ as $J/t$ decreases.
This is naturally understood since the energy to excite the system
to higher spin state becomes less as $J/t$ decreases.
In the supersymmetric case, the GWF is in good agreement with the
BA result also for the quarter filling.\refto{Quaisser}
Furthermore, the non-interacting result agrees well for all
the values of $H/t$.
\par

\par\medskip
\noindent{\it 6.3\quad Comparisons and Discussions}
\par\smallskip\noindent
First, we compare the above result with the Hubbard model.
Exact results\refto{Takahashi,Shiba,Kawakami3} show that in the
limit of $n\rightarrow 1$,
$\chi_{\rm c}$ diverges as $\chi_{\rm c}=\alpha/\delta$,
where $\alpha$ is a numeral factor which depends on $U/t$.\refto{Kawakami3}
The value of $\alpha$ changes from zero for $U/t=0$ to $1/2\pi^2$
for $U/t=\infty$, which is the same with the $t$-$J$ model with
$J/t=0$.
On the other hand, $\chi_{\rm s}$ converges upon a finite value
as $n$ approaches 1.\refto{Takahashi}
In the strong-$U$ limit the expansion coefficient in $t^2/U$ of
$\chi_{\rm s}$ is the same as the $t$-$J$ model.
These features qualitatively agree with the variational results
of the present study as well as the exact one for the $t$-$J$ model.
\par
Meanwhile, Otsuka\refto{Otsuka} and Furukawa and Imada\refto{FI}
investigated the critical
behavior of $\chi_{\rm c}$ and $\chi_{\rm s}$ for the Hubbard model
on 2D square lattice by using quantum Monte Carlo methods.
According to their results, essentially the same properties with
the 1D models are observed.
Thus one can consider that a universal profile of a kind of Mott
transition appears in these results of $\chi_{\rm c}$ and
$\chi_{\rm s}$.
\par
Keeping these in mind, next let us compare with the Brinkman-Rice
transition in the Hubbard model.
As summarized in Appendix B, $\chi_{\rm s}$ calculated with Gutzwiller
approximation diverges when $U\rightarrow U_{\rm c}\ (n=1)$ or
$n\rightarrow 1$ $(U\ge U_{\rm c})$.
On the other hand, $\chi_{\rm c}$ remains finite for a finite
value of $U$, even if $U>U_{\rm c}$, although effective mass $m^*/m$
diverges similarly as $\chi_{\rm s}$.
This indicates that the Brinkman-Rice transition describes a quite
different type of metal-insulator transition from those in the
1D and 2D Hubbard models and the 1D $t$-$J$ model.
\par

Lastly we mention the results in the long-range
$t$-$J$ model.\refto{Kura}
In this model, susceptibilities are written as
$$
\chi_{\rm c}^{-1}={{\pi^2}\over{2}}(1-n),\qquad\qquad
\chi_{\rm s}^{-1}={{\pi^2}\over{2}}(1-m). \eqno(6.6)
$$
Notice that $\chi_{\rm s}$ does not depend on $n$ and $\chi_{\rm c}$
does not on $m$, since the contributions to energy of $n$ and $m$ are
mutually independent, and that there are no system size
dependence, because the size-dependent terms in energy are
exhausted in the linear order.
The value of $\chi_{\rm s}$ is constant ($2/\pi^2$) under zero
field irrespective of $n$, which is the same for the
nearest-neighbor Heisenberg model and is close to the variational
value (GWF).
The divergent behavior of $\chi_{\rm c}$ near the half filling is
also similar to the $t$-$J$ model (see Table IV).
On the other hand, a quite different feature appear in the low
electron density region, where there is no divergence.
This difference is originated in the band structure of the model;
the non-interacting long-range model has linear energy dispersion
and there is no divergence in density of state at the band edge.
 From this we can see that electron correlation affects severely
high-electron-density regime, while in the low-density regime
the density of states of the original non-interacting system
determines the charge susceptibility.
\par
In Fig.\ 24 we also plot the magnetization curve
of the long-range $t$-$J$ model:\refto{Kawakami2,YK}
$$
m=1-2\sqrt{{{1}\over{4}}-{{H}\over{\pi^2t}}}. \eqno(6.7)
$$
This formula does not depend on $n$.
The critical field $H_{\rm s}$ for the long-range model is
somewhat larger ($\pi^2/4$ for $n=1$ and $3\pi^2/16$ for $n=0.5$)
than the value of the ordinary $t$-$J$ model.
And the difference is larger for $n=0.5$.
This is probably because the long-range exchange terms
tend to disturb ferromagnetic spin alignment.
Especially in the low density region, where the particle distance
is large, the long-range terms play important roles.
\par
\fontA\smallskip
\par\noindent
\fontB\noindent
\beginsection \S7. Summary and Discussion
\par
\fontA\smallskip
\par\noindent
We have pursued the ground-state properties of the
one-dimensional $t$-$J$ model in the light of wave functions.
By comparing the variational Monte Carlo results with those of
the exact diagonalization, the Bethe Ansatz and the Gutzwiller
approximation, we have obtained some remarkable aspects
as follows:
\par\noindent
[1] From diagonalization, exact energy, momentum distribution,
and spin and charge correlation functions are obtained,
which show unusual behaviors as $J/t$ increases.
\par\noindent
[2] In the supersymmetric case ($J/t=2$) the Gutzwiller wave
function is an extremely good state for bulk quantities.
In the low-density limit the GWF becomes exact.
\par\noindent
[3] Whole parameter space spanned by electron density $n$ and
coupling strength $J/t$ is well described by the repulsive or
the attractive Jastrow-type wave functions.
The parameter space can be roughly divided into:

(a)\qquad $J/t<2$ \qquad repulsive region (RJWF),

(b)\qquad $J/t=2$ \qquad free-like region (GWF),

(c)\qquad $J/t>2$ \qquad attractive region (AJWF).

\par\noindent
In the region (a), the electron hopping term is dominant.
In the region (c), dominant is the exchange term, which
induces attractive interaction.
In the supersymmetric case (b), the two terms are well balanced and
a kind of ``non-interacting" state is realized, especially
for the low electron density.
\par\noindent
[4] Phase transition from a homogeneous state to a phase-separated
state is quantitatively described within the AJWF.
\par\noindent
[5] In the Jastrow wave functions, a short-range part of intersite
correlation factors are responsible for relatively high energy
processes which determine bulk properties like energy and magnitude
of correlation functions.
On the other hand, a long-range part is mainly concerned in low
energy processes near the Fermi surface, which cause the critical
properties characteristic of the Tomonaga-Luttinger liquid.
\par\noindent
[6] The Jastrow wave functions reproduces charge and spin
susceptibilities and magnetization curve correctly, in contrast
with the Gutzwiller approximation.
\par
Keeping these results in mind, we mention some remaining issues.
\par
In the region of low electron density and $J/t>2$, there exists
a spin gap state (without a charge gap).\refto{Ogata,Hellberg3}
We have not found an indication of a spin gap in the trial
functions used.
On the other hand, Chen and Lee\refto{Chen} introduced a trial
state for a gas of singlet pairs, and showed that there is
a region where this function is stabler than the TL-liquid wave
function (5.1).
\par
An interesting extension of the present method is to 2D systems.
In 2D we do not know even the ground state---the Fermi liquid
or the TL liquid for the metallic regime.
Furthermore, magnetically ordered phases can be stabilized,
near the half filling.\refto{VM2}
Actually the TL-liquid wave function has been extended to a 2D system
by Valenti and Gros.\refto{Valenti}
According to their results, the energy lowering by their function
is very small (1\%) compared with the simple Gutzwiller wave
function.
Also the critical exponent at $k_{\rm F}$ in the momentum
distribution is small, assuming that it exists.
Since critical properties are in a low energy scale, it may not
be easy to judge the realization of a TL-liquid state in 2D
only by the stability in energy.
\par
Recently the behaviors of specific-heat coefficient,
effective mass, $\chi_{\rm c}$, $\chi_{\rm s}$, etc.\
are investigated experimentally for the high-$T_{\rm c}$
superconductors and related Mott insulators,\refto{Tokura}
in connection with the metal-insulator transition.
The results of these experiments together with the
theoretical calculations\refto{Otsuka,FI} have suggested
reconsideration of the appropriate Hamiltonian,
namely whether the simple $t$-$J$ model is pertinent
to describe various aspects of the high-$T_{\rm c}$ cuprates.
\par
In these contexts, 2D systems have to be further studied
with the variation theory.
\par
\medskip\fontB
\centerline{Acknowledgment}
\smallskip\fontA
The authors are grateful to T.M.\ Rice, H.\ Shiba, Y.\ Kuramoto,
D.\ Vollhardt, W.O.\ Putikka, M.U.\ Luchini, S.\ Sorella and
P.-A.\ Bares for fruitful discussions.
One of the authors (H.Y.) thanks N.\ Kawakami, W.\ Metzner,
Y.\ \=Ono, T.\ Pruschke, K.\ Sano, B.S.\ Shastry, P.G.J.\ van Dongen
and F.C.\ Zhang for useful comments.
This work is supported partly by Grant-in-Aids for Scientific Research
on Priority Areas, ``Computational Physics as a New Frontier
in Condensed Matter Research" and ``Science of High-$T_{\rm c}$
Superconductivity" and for Encouragement of Young Scientists given
by Ministry of Education, Science and Culture, Japan.
\par
\fontB\noindent
\beginsection Appendix A: Analytical Approach to the Gutzwiller
Wave Function.
\par
\fontA\smallskip
\par\noindent
In this appendix, we summarize the analytic expressions for various
expectation values in the GWF.  They were developed by Metzner and
Vollhardt,\refto{Voll1} and by Gebhard and Vollhardt\refto{Voll2}
for the Hubbard model.
In order to apply to the 1D $t$-$J$ model, we carry out the
calculation of $\langle n_i n_j \rangle$ and $\langle S^z_i S^z_j \rangle$.

The momentum distribution for spin $\sigma$ is obtained
by an infinite summation\refto{Voll1}
$$
\eqalignno{
n_\sigma(k)= &n^0_\sigma(k) -(1-g)^2n_{-\sigma}n^0_\sigma(k) \cr
&+{1\over (1+g)^2}\sum_{m=2}^\infty (g^2-1)^m \{1-(1-g^2)n^0_\sigma(k)\}
f_{m\sigma}(k),
&{({\rm A}.1)}
}$$
where $g$ is identical to $\eta(0)$ in our notation (eq.\ (4.1)), and
thus $g=0$ for the $t$-$J$ model.
$n^0_\sigma(k)$ is the momentum distribution of the non-interacting
system given by $n^0_\sigma(k) = \theta(k_{\rm F}-|k|)$, and
$n_\sigma = n/2$.
In 1D case, $f_m(k)$ is given by a polynomial of order $\leq m$.
According to the notation in Ref.\refto{Voll1}, we can summarize
as follows:
$$
f_m(k)=\cases{
   n^m R_m(k)\ , \cr
         \quad\qquad({\rm in\ region\ (I)}: 0\leq k\leq k_{\rm F}) \cr
         \ \cr
   n^m Q_m(k) + C_{m-1}\ , \cr
         \quad\qquad({\rm in\ region\ (II)}: k_{\rm F}\leq k\leq
           \min (3k_{\rm F}, 2\pi-3k_{\rm F})  \cr
         \ \cr
   C_{m-1}\ , \cr
         \quad\qquad({\rm in\ region\ (III)}: 3k_{\rm F}\leq k\leq \pi,
           {\rm when}\ 3k_{\rm F}<\pi)  \cr
         \ \cr
   n^m (Q_m(k) + Q_m(2\pi-k)) + C_{m-1}\ , \cr
         \quad\qquad({\rm in\ region\ (IV)}: 2\pi-3k_{\rm F}\leq k\leq \pi,
           {\rm when}\ 2\pi-3k_{\rm F}<\pi)  \cr
}\eqno{({\rm A}.2)}
$$
where
$$
C_{m-1} = (-1)^m {n^m\over 2m},
$$
$$
R_m(k)=\sum_{j=1}^m {R_m^{(j)} \over j!}
\bigl({k\over 2\pi n}-{1\over 4}\bigr)^j,
$$
$$
Q_m(k)=\sum_{j=1}^m {Q_m^{(j)} \over j!}
\bigl({k\over 2\pi n}-{3\over 4}\bigr)^j.
\eqno{({\rm A}.3)}
$$
The coefficients of these Taylor series are determined via a
recursion relation;
$$
2(m-j+1)R_{m+1}^{(j)} = -(2m-2j+1)R_{m}^{(j)} - Q_{m+1}^{(j+1)},
$$
$$
Q_{m+1}^{(j+2)} = 2(m-2j)Q_{m+1}^{(j+1)} + 2mR_{m}^{(j)}
                - 4j(m-j+1)(R_{m}^{(j)}+R_{m+1}^{(j)}),
$$
$$
R_{m+1}^{(m+1)}=\cases{0 &\qquad $m+1 = $ odd \cr
                       -2Q_{m+1}^{(m+1)}, &\qquad $m+1 = $ even \cr}
\eqno{({\rm A}.4)}
$$
with initial values
$$
R_{1}^{(j)}=-{1\over 2}\delta_{j,0},
$$
$$
Q_{m+1}^{(0)}=Q_{m+1}^{(1)}=0.
$$

Apparently the series in (A.1) gives an expansion with respect to
$n$ in the low density region.  For $m=2$, we get
$$
f_2(k)=\cases{
   n^2\bigl( {k^2\over 4\pi^2n^2}+{5\over 16}\bigr) &\qquad in (I)\cr
   -{n^2\over 2}\bigl( {k\over 2\pi n}-{3\over 4}\bigr)^2+{n^2\over 4}
                                                    &\qquad in (II)\cr
   {n^2\over 4}                                     &\qquad in (III)\cr
   -{n^2\over 2}\bigl( {k\over 2\pi n}-{3\over 4}\bigr)^2
   -{n^2\over 2}\bigl( {2\pi -k\over 2\pi n}-{3\over 4}\bigr)^2+{n^2\over 4}
                                                    &\qquad in (IV)\cr
}\eqno{({\rm A}.5)}
$$
Then the kinetic energy in the dilute limit is
$$
E_t = -2t\biggl( n-{n^2 \over 2}-{\pi^2 \over 24}n^3-{n^3 \over 3}
\biggr) +O(n^4).
\eqno{({\rm A}.6)}
$$

Compact analytic expressions for $S(k)$ and $N(k)$ are given in
(3.17) and (3.18).
Their Fourier transforms give the spin and charge correlation
functions in real space.  Through a straightforward calculation, we get
$$
4\langle S^z_i S^z_{i+r}\rangle  = {(-1)^r\over \pi r}
\bigl\{ {\rm Si}(\pi r)-{\rm Si}\bigl((1-n)\pi r \bigr)\bigr\},
\eqno{({\rm A}.7)}
$$
and
$$
\eqalign{
\langle n_i n_{i+r} \rangle &=
n^2+{1\over 2\pi^2 r^2}\bigl(\cos 4k_{\rm F}r -1\bigr) \cr
&-{(-1)^r\over \pi r}
\bigl\{ {\sin n\pi r \over \pi r}+(1-n)\cos n\pi r \bigr)
\bigl\{ {\rm Si}(\pi r)-{\rm Si}\bigl((1-n)\pi r\bigr)\bigr\},
}\eqno{({\rm A}.8)}
$$
where $r$ is an integer, $r\geq 1$.  When $r=1$, (A.7) and (A.8)
give the exchange energy given in (3.3).  In the low density limit,
we obtain
$$
\langle S^z_i S^z_{i+1} \rangle = -{n^2 \over 8}-{n^3 \over 12}+O(n^4),
$$
and
$$
\langle n_i n_{i+1}\rangle  = {n^2 \over 2}+{n^3 \over 3}+O(n^4).
\eqno{({\rm A}.9)}
$$
Total energy is
$$
E = -2tn +\bigl(t-{J\over 2}\bigr)n^2+ {\pi^2 \over 12}tn^3
    +\bigl({2 \over 3}t-{1 \over 3}J\bigr)n^3 +O(n^4).
\eqno{({\rm A}.10)}
$$
At $J/t=2$, we get (3.12).

\fontB\noindent
\beginsection Appendix B: $\chi_{\rm s}$ and $\chi_{\rm c}$ around
Brinkman-Rice Transition.
\par
\fontA\smallskip
\par\noindent
Brinkman and Rice discussed the metal-insulator transition for the
Hubbard model based on the Gutzwiller approximation (GA).\refto{Brinkman}
Although it was confirmed by solving the variation problem accurately
that this transition does not exist in the realistic (1,2 and 3)
dimensions,\refto{YS1,Voll1} the conception of the Brinkman-Rice
transition is still widely used.
In this Appendix we briefly review the behavior of $\chi_{\rm s}$
and $\chi_{\rm c}$ in the GA, especially around the Brinkman-Rice
transition for the comparison in \S6.\refto{Voll3}
\par
According to the GA, the state of $n\ne 1$ in the Hubbard model is always
metallic and has a Fermi surface with a finite discontinuity $q$ of
$n(k)$ at $k=k_{\rm F}$.
On the other hand, in the half filling the GA gives a metallic state
for $U<U_{\rm c}$ and an insulating one for $U>U_{\rm c}$,
where $U_{\rm c}=8|\varepsilon_0|$ and $\varepsilon_0$ is the
energy of the non-interacting system; the Brinkman-Rice transition
occurs at $U=U_{\rm c}$.
When $U$ increases from under $U_{\rm c}$ fixing $n=1$,
$\chi_{\rm s}$ as well as effective mass $m^*/m(\propto q^{-1})$
diverges as $1/(1-(U/U_c)^2)$.
$\chi_{\rm s}$ and $m^*/m$ remain infinite for $U\ge U_{\rm c}$.
Meanwhile, $\chi_{\rm c}$ decreases with increasing $U$ and
vanish at $U=U_{\rm c}$ and then increases for $U>U_{\rm c}$.
In Fig.\ 25 we actually plot the numerical GA values of $\chi_{\rm s}$
and $\chi_{\rm c}$.
According to the Fermi liquid theory, charge susceptibility is
related to $m^*/m$ as
$$
\chi_{\rm c}={{m^*/m}\over{1+F^{\rm s}_0}}\chi^0_{\rm c}, \eqno({\rm B}.1)
$$
where $F^{\rm s}_0$ is the usual Landau parameter and
$\chi^0_{\rm c}$ is the value for the non-interacting case.
Therefore $F^{\rm s}_0$ is more divergent than $m^*/m$
at $U=U_{\rm c}$.
\par\bigskip\bigskip
\references

\par\noindent
\item{*} E-mail address: yoko@cmpt01.phys.tohoku.ac.jp
\par\noindent
\item{**} E-mail address: ogata@sola.c.u-tokyo.ac.jp
\par

\refis{Anderson1} P.\ W.\ Anderson: \journal{Science}{235}{1196}{1987};
     F.\ C.\ Zhang and T.\ M.\ Rice: \journal{\PRB}{37}{3759}{1988}.

\refis{Gros} C.\ Gros, R.\ Joynt and T.M.\ Rice: \journal{\PRB}{36}{381}{1987}.

\refis{LiebWu} E.H.\ Lieb and F.Y.\ Wu: \journal{\PRL}{20}{1445}{1968}.

\refis{Haldane} F.D.M.\ Haldane: \journal{\PRL}{45}{1358}{1980};
                      \journal{\JPC}{14}{2585}{1981}.

\refis{Schulz} H.J.\ Schulz: \journal{\PRL}{64}{2381}{1990}.

\refis{FK} H.\ Frahm and V.E.\ Korepin: \journal{\PRB}{42}{10553}{1990}.

\refis{KY1}  N.\ Kawakami and S.-K.\ Yang: \journal{\PLA}{148}{359}{1990}.

\refis{KY2} N.\ Kawakami and S.-K.\ Yang: \journal{\PRL}{65}{2309}{1990};
        \journal{J.\ Phys.\ Cond.\ Mat.}{3}{5983}{1991}.

\refis{YS1} H.\ Yokoyama and H.\ Shiba: \journal{\JPSJ}{56}{1490}{1987}.

\refis{VM2} H.\ Yokoyama and H.\ Shiba: \journal{\JPSJ}{56}{3570}{1987},
{\it ibid.}, {\bf 57}, 2482 (1988); C.\ Gros: \journal{\PRB}{38}{931}{1988},
\journal{Ann.\ Phys}{189}{53}{1989}; T.\ Giamarchi and C.\ Lhuillier:
\journal{\PRB}{43}{12943}{1991}.

\refis{Voll1} W.\ Metzner and D.\ Vollhardt: \journal{\PRB}{37}{7382}{1988}.

\refis{Voll2} F.\ Gebhard and D.\ Vollhardt: \journal{\PRB}{38}{6911}{1988}.

\refis{Gutzwiller} M.C.\ Gutzwiller: \journal{\PRL}{10}{159}{1963};
          \journal{\PR}{134}{A1726}{1965}.

\refis{YS2} H.\ Yokoyama and H.\ Shiba: \journal{\JPSJ}{59}{3669}{1990}.

\refis{Imada} M.\ Imada: \journal{\JPSJ}{59}{4121}{1990}; in
     {\it Quantum Simulations of Condensed Matter Phenomena}, \rm ed.\
     J.\ D.\ Dell and J.\ E.\ Gubernatis (World Scientific, Singapore,
     1990), p.\ 127.

\refis{YO} H.\ Yokoyama and M.\ Ogata: \journal{\PRL}{67}{3610}{1991}.

\refis{Hellberg1} C.\ S.\ Hellberg and E.\ J.\ Mele:
     \journal{\PRB}{44}{1360}{1991}; \journal{\IJMP}{5}{1791}{1991}.

\refis{Hellberg2} C.\ S.\ Hellberg and E.\ J.\ Mele:
     \journal{\PRL}{67}{2080}{1991}.

\refis{GL} T.\ Giamarchi and C.\ Lhuillier: \journal{\PRB}{43}{12943}{1991}.

\refis{Valenti} R.\ Valenti and C.\ Gros: \journal{\PRL}{68}{2402}{1992};
C.\ Gros and R.\ Valenti: \journal{Mod.\ Phys.\ Lett.\ B}{7}{119}{1992}.

\refis{KH} N.\ Kawakami and P.\ Horsch: \journal{\PRL}{68}{3110}{1992};
     C.\ S.\ Hellberg and E.\ J.\ Mele: \journal{\PRL}{68}{3111}{1992}.

\refis{tjBA} B.\ Sutherland: \journal{\PRB}{12}{3795}{1975};
     P.\ Schlottmann: \journal{\PRB}{36}{5177}{1987}.

\refis{tjBA2} P.-A.\ Bares and G.\ Blatter:
     \journal{\PRL}{64}{2567}{1990};
     P.-A.\ Bares, G.\ Blatter and M.\ Ogata:
     \journal{\PRB}{44}{130}{1991}.

\refis{Ogata} M.\ Ogata, M.\ U.\ Luchini, S.\ Sorella and F.\ F.\ Assaad:
     \journal{\PRL}{66}{2388}{1991}.

\refis{Assaad} F.\ F.\ Assaad and D.\ W\"urtz: \journal{\PRB}{44}{2681}{1991}.

\refis{OgataShiba} M.\ Ogata and H.\ Shiba: \journal{\PRB}{41}{2326}{1990};
     H.\ Shiba and M.\ Ogata:
       \journal{Int.\ J.\ of Mod.\ Phys.\ }{B5}{31}{1991}.

\refis{footnote1}
In this case, level crossings take place and thus the singlet
state does not have the lowest energy in the small-$J$ region.
However we identify the energy level corresponding to the lowest singlet
state by following the singlet state as a function of $J$.

\refis{footnote2}
Other combinations of the powers of $1/N_{\rm a}$ are tried to show that
this formula gives the best fitting as expected.

\refis{footnote3}
There was an unfortunate typographical error in eq.\ (2) of the
previous report.\refto{YO}
The present equation (4.1) is the correct one.

\refis{footnote4}
If one introduces spin-dependent correlation factor keeping
the singlet nature, one has to use spin operators isotropically,
which means that the Jastrow factor includes off-diagonal elements.
This issue is left for future study.

\refis{Kura} Y.\ Kuramoto and H.\ Yokoyama: \journal{\PRL}{67}{1338}{1991}.

\refis{Kato} Y.\ Kato and Y.\ Kuramoto: \journal{\PRL}{74}{1222}{1995}.

\refis{YK} H.\ Yokoyama and Y.\ Kuramoto: \journal{\JPSJ}{61}{3046}{1992}.

\refis{YKO} H.\ Yokoyama, Y.\ Kuramoto and M.\ Ogata:
to be published in J.\ Phys.\ Soc.\ Jpn.\ Suppl.\ (1995).

\refis{Ha} Z.N.C.\ Ha and F.D.M.\ Haldane: \journal{\PRB}{46}{9359}{1992}.

\refis{Nishimori} T.\ Oguchi, H.\ Nishimori, and Y.\ Taguchi:
\journal{\JPSJ}{55}{323}{1986}.

\refis{AndersonSH} P.\ W.\ Anderson, B.\ S.\ Shastry, and D.\ Hristopulos,
\journal{\PRB}{40}{8939}{1989}.

\refis{SP} T.\ Pruschke and H.\ Shiba: \journal{\PRB}{44}{205}{1991};
\journal{\PRB}{46}{356}{1992}.

\refis{Griffiths} R.\ B.\ Griffiths: \journal{\PR}{133}{A786}{1964}.

\refis{Quaisser} M.\ Quaisser, A.\ Schadschneider and J.\ Zittartz:
J. Phys. A {\bf 25}, L1127 (1992).

\refis{Shiba} H.\ Shiba: \journal{\PRB}{6}{930}{1972}.

\refis{Kawakami3} T.\ Usuki, N.\ Kawakami and A.\ Okiji:
\journal{\PLA}{135}{476}{1989};
N.\ Kawakami and A.\ Okiji: {\it Strong Correlation and Superconductivity},
ed.\ H.\ Fukuyama, S.\ Maekawa and A.\ P.\ Malozemoff (Springer Verlag, 1989).

\refis{Otsuka} H.\ Otsuka: \journal{\JPSJ}{59}{2916}{1990}.

\refis{FI} N.\ Furukawa and M.\ Imada: \journal{\JPSJ}{61}{3331}{1992},
\journal{\JPSJ}{62}{2557}{1993}.

\refis{footnote5} For the Hubbard model with finite $U/t$ in the
half-filled band, the behavior of the GWF is different.
As $H$ approaches $H_{\rm s}$, total energy
comes to have two local minima, one of which is at $m=n$.
And the global minimum switches from one at $0<m<n$ to the other
at $m=n$ abruptly.
Thus $m$ has a jump to the full momentum at the critical
field.
This aspect is common with the Gutzwiller approximation\refto{Voll3}
but not consistent with the exact result.\refto{Takahashi}

\refis{Kawakami2} N.\ Kawakami: \journal{\PRB}{45}{7525}{1992};
          \journal{\PRB}{46}{1005}{1992}.

\refis{Voll3} Detailed discussions on the Gutzwiller approximation
are given in, D.\ Vollhardt, \journal{\RMP}{56}{99}{1984}.

\refis{Brinkman} W.\ F.\ Brinkman and T.\ M.\ Rice,
\journal{\PRB}{2}{4302}{1970}.

\refis{OgataKoba} M.\ Ogata: {\it Correlation Effects in Low-Dimensional
Electron Systems}, ed.\ A.\ Okiji and N.\ Kawakami (Springer Verlag 1994)
p.121; K.\ Kobayashi: unpublished.

\refis{Chen} Y.C.\ Chen and T.K.\ Lee: \journal{\PRB}{47}{11548}{1993}.

\refis{Koba} K.\ Kobayashi and K.\ Iguchi: \journal{\PRB}{47}{1775}{1993}.

\refis{Hellberg3} C.S.\ Hellberg and E.J.\ Mele: \journal{\PRB}{48}{646}{1993};
\journal{Physica B}{199 \& 200}{322}{1994}.

\refis{Heeb} E.S.\ Heeb and T.M.\ Rice: preprint.

\refis{Tokura} For instance,
Y.\ Tokura et al.: \journal{\PRL}{70}{2126}{1993};
K.\ Kumagai et al.: \journal{\PRB}{48}{7636}{1993};
N.E.\ Phillips et al.: {\it Progress in Low Temperature Physics},
ed. D.\ Brewer (Elsevier, the Netherland, 1990).

\refis{Takahashi} M.\ Takahashi: \journal{\PTP}{42}{1092}{1969}.

\endreferences
\vfil\eject

\TAB Ground-state energies of the 1D $t$-$J$ model for the quarter-filled
case ($n=1/2$).  They are extrapolated to the thermodynamic limit
using the formula (2.6).
$\epsilon_\infty$ and $\epsilon'_\infty$ are obtained from the two set
of boundary conditions (see the text).
The unit of the energy is $t$.

\par
$$
\vbox{\tabskip=0pt \offinterlineskip
\def\tablerule{\noalign{\hrule}}
\halign to 180 pt {\strut# \tabskip=1em plus 1em &
  \hfil#\hfil& \vrule#&\hfil#\hfil&
  \hfil#\hfil \tabskip=0pt\cr\tablerule
&\omit\hidewidth $J/t$ \hidewidth&&
 \omit\hidewidth $\epsilon_\infty$  \hidewidth&
 \omit\hidewidth $\epsilon'_\infty$ \hidewidth\cr\tablerule
\tablerule
& 0.0 &&  $-0.6366197$
       &  $-0.6366198$ \cr\tablerule
& 0.2 &&  $-0.6578750$
       &  $-0.6575043$ \cr\tablerule
& 0.4 &&  $-0.6804029$
       &  $-0.6806269$ \cr\tablerule
& 0.6 &&  $-0.7041767$
       &  $-0.7041768$ \cr\tablerule
& 0.8 &&  $-0.7291701$
       &  $-0.7291701$ \cr\tablerule
& 1.0 &&  $-0.7553587$
       &  $-0.7553586$ \cr\tablerule
& 1.2 &&  $-0.7827211$
       &  $-0.7827209$ \cr\tablerule
& 1.4 &&  $-0.8112399$
       &  $-0.8112396$ \cr\tablerule
& 1.6 &&  $-0.8409031$
       &  $-0.8409028$ \cr\tablerule
& 1.8 &&  $-0.8717049$
       &  $-0.8717046$ \cr\tablerule
& 2.0 &&  $-0.9036477$
       &  $-0.9036477$ \cr\tablerule
& 2.2 &&  $-0.9367449$
       &  $-0.9367452$ \cr\tablerule
& 2.4 &&  $-0.9710250$
       &  $-0.9710259$ \cr\tablerule
& 2.6 &&  $-1.0065398$
       &  $-1.0065415$ \cr\tablerule
& 2.8 &&  $-1.0433801$
       &  $-1.0433827$ \cr\tablerule
& 3.0 &&  $-1.0817130$
       &  $-1.0817179$ \cr\tablerule
& 3.2 &&  $-1.1219079$
       &  $-1.1219533$ \cr\tablerule
}}$$

\vfil\eject

\TAB Ground-state energies of the 1D $t$-$J$ model for $n=3/4$
similarly obtained as in the quarter-filled case in Table I.

\par
$$
\vbox{\tabskip=0pt \offinterlineskip
\def\tablerule{\noalign{\hrule}}
\halign to 180 pt {\strut# \tabskip=1em plus 1em &
  \hfil#\hfil& \vrule#&\hfil#\hfil&
  \hfil#\hfil \tabskip=0pt\cr\tablerule
&\omit\hidewidth $J/t$ \hidewidth&&
 \omit\hidewidth $\epsilon_\infty$  \hidewidth&
 \omit\hidewidth $\epsilon'_\infty$ \hidewidth\cr\tablerule
\tablerule
& 0.0 &&  $-0.4501050$
       &  $-0.4502187$ \cr\tablerule
& 0.2 &&  $-0.5216656$
       &  $-0.5219004$ \cr\tablerule
& 0.4 &&  $-0.5943201$
       &  $-0.5944475$ \cr\tablerule
& 0.6 &&  $-0.6678961$
       &  $-0.6680324$ \cr\tablerule
& 0.8 &&  $-0.7422736$
       &  $-0.7424195$ \cr\tablerule
& 1.0 &&  $-0.8173650$
       &  $-0.8175213$ \cr\tablerule
& 1.2 &&  $-0.8931052$
       &  $-0.8932726$ \cr\tablerule
& 1.4 &&  $-0.9694457$
       &  $-0.9696252$ \cr\tablerule
& 1.6 &&  $-1.0463508$
       &  $-1.0465437$ \cr\tablerule
& 1.8 &&  $-1.1237962$
       &  $-1.1240038$ \cr\tablerule
& 2.0 &&  $-1.2017671$
       &  $-1.2019909$ \cr\tablerule
& 2.2 &&  $-1.2802580$
       &  $-1.2805002$ \cr\tablerule
& 2.4 &&  $-1.3592732$
       &  $-1.3595368$ \cr\tablerule
& 2.6 &&  $-1.4388286$
       &  $-1.4391186$ \cr\tablerule
& 2.8 &&  $-1.5189551$
       &  $-1.5192801$ \cr\tablerule
& 3.0 &&  $-1.5997064$
       &  $-1.6000835$ \cr\tablerule
& 3.2 &&  $-1.6811763$
       &  $-1.6816414$ \cr\tablerule
& 3.4 &&  $-1.7635421$
       &  $-1.7641837$ \cr\tablerule
}}$$
\par
\vfill\eject
\TAB Comparison of expectation values for four kinds of
correlation factors hybridized
(a) between the RJWF eq.(4.2) ($\zeta=6.9847$) and eq.(5.1) ($\nu=0.5$)
for the repulsive case,
and (b) between the AJWF eq.(4.2) ($\alpha=0.6804$, $\beta=0.5481$)
and eq.(5.1) ($\nu=-0.15$) for the attractive case.
Digits include some statistical errors.
\par
$$
\vbox{\tabskip=0pt \offinterlineskip
\def\tablerule{\noalign{\hrule}}
\def\d#1{\smash{\lower1ex\hbox{#1}}}
    \halign {\strut#& \vrule# \tabskip=1em plus2em&
         \hfil#\hfil& \vrule#& \hfil#\hfil& \vrule#&
         \hfil#\hfil& \vrule#& \hfil#\hfil& \vrule#&
         \hfil#\hfil& \vrule#& \hfil#\hfil& \vrule#
         \tabskip=0pt\cr\tablerule
   &&(a)&&$E_t/t$&&$E_J/J$&&$n(k=0)$&
             &$\delta n(k)\ <k_{\rm F}$&&$\delta n(k)\ >k_{\rm F}$&\cr
   \tablerule
   &&case 1)&&$-0.60865$&&$-0.13972$&&$0.73826$&&$-0.02925$&&$-0.02948$&\cr
   &&case 2)&&$-0.60959$&&$-0.12047$&&$0.70705$&&$-0.02665$&&$-0.02811$&\cr
   &&case 3)&&$-0.60880$&&$-0.13955$&&$0.73804$&&$-0.01958$&&$-0.01989$&\cr
   &&case 4)&&$-0.60983$&&$-0.12037$&&$0.70816$&&$-0.02242$&&$-0.02178$&\cr
   \tablerule
    }}
$$
\par
$$
\vbox{\tabskip=0pt \offinterlineskip
\def\tablerule{\noalign{\hrule}}
\def\d#1{\smash{\lower1ex\hbox{#1}}}
    \halign {\strut#& \vrule# \tabskip=1em plus2em&
         \hfil#\hfil& \vrule#& \hfil#\hfil& \vrule#&
         \hfil#\hfil& \vrule#& \hfil#\hfil& \vrule#&
         \hfil#\hfil& \vrule#& \hfil#\hfil& \vrule#
         \tabskip=0pt\cr\tablerule
   &&(b)&&$E_t/t$&&$E_J/J$&&$n(k=0)$&
             &$\delta n(k)\ <k_{\rm F}$&&$\delta n(k)\ >k_{\rm F}$&\cr
   \tablerule
   &&case 1)&&$-0.55403$&&$-0.17331$&&$0.74876$&&$-0.00589$&&$-0.01103$&\cr
   &&case 2)&&$-0.54695$&&$-0.17619$&&$0.74777$&&$-0.00309$&&$-0.00559$&\cr
   &&case 3)&&$-0.55392$&&$-0.17334$&&$0.74871$&&$-0.00096$&&$-0.00553$&\cr
   &&case 4)&&$-0.54701$&&$-0.17615$&&$0.74745$&&$-0.00090$&&$-0.00279$&\cr
   \tablerule
   }}
$$
\par\bigskip\bigskip
\TAB Coefficient of $1/n$ and $1/\delta$ of charge
susceptibility in the limit of $n\rightarrow 0$ and
$\delta\rightarrow 0$, respectively.
For comparison we also show the values of the free electron system
and the supersymmetric $t$-$J$ model with long-range exchange and
transfer, denoted by `Free' and `L.-r.\ $t$-$J$' respectively in the
first column.

$$
\vbox{\tabskip=0pt \offinterlineskip
\def\tablerule{\noalign{\hrule}}
\def\d#1{\smash{\lower1ex\hbox{#1}}}
    \halign {\strut#& \vrule# \tabskip=1em plus2em&
         \hfil#\hfil& \vrule#& \hfil#\hfil& \vrule#&
         \hfil#\hfil& \vrule#& \hfil#\hfil& \vrule#&
         \hfil#\hfil& \vrule# \tabskip=0pt\cr\tablerule
   &&&&\multispan3\hfil $n\rightarrow 0$\hfil
          &&\multispan3\hfil $\delta\rightarrow 0$\hfil&\cr\tablerule
   &&$J/t$&&Variational&&Exact&&Variational&&Exact&\cr\tablerule
   &&0&&0.046&&${{1}\over{2\pi^2}}=0.05066$
      &&0.054&&${{1}\over{2\pi^2}}=0.05066$&\cr
   &&1&&0.053&&{\bf ---}
      &&0.062&&{\bf ---}&\cr
   &&2&&0.204&&${{2}\over{\pi^2}}=0.20264$
      &&0.070&&${{16(\ln 2)^2}\over{3\pi^2\zeta(3)}}=0.21598$&\cr
   \tablerule
   &&Free&&{\bf ---}&&${{2}\over{\pi^2}}=0.20264$
         &&{\bf ---}&&Finite $\chi_{\rm c}$&\cr
   \tablerule
   &&L.-r.\ $t$-$J$&&Exact&&Finite $\chi_{\rm c}$
         &&Exact&&${{2}\over{\pi^2}}=0.20264$&\cr
   \tablerule
    }}
$$

\vfil\eject

\beginsection \centerline{Figure Captions}
\global\FGnum=0

\FIG
(a) The momentum distribution function, (b) spin correlation function,
and (c) charge correlation function
for the quarter-filled case at
$J/t=0.5$, 1.0, 2.0 and 3.0 obtained in the exact diagonalization of
the systems with 4, 8, 12 and 16 sites.
For $J=0$, we show the result in
the wave function for the $U\to\infty$ Hubbard model with 52
sites.\refto{OgataShiba,SP}

\FIG
Ground-state energy of the 1D $t$-$J$ model
as a function of $J/t$ at $n=0.5$.
Extrapolated values ($N_{\rm a}\to\infty$) of the exact diagonalization,
the energy of the fully phase separated (Heisenberg) state, and
variational energies of the three types of variational functions are
compared.
The transition to a phase separated state occurs at $J/t\sim 3.3$
shown by a hollow arrow.
In the VMC calculations systems with 50-100 sites are used.

\FIG
Comparison of the total energy per site
between the Gutzwiller wave function (closed circles)
and Bethe Ansatz (Solid line) for $J/t=2$ as a function of electron
density.
Although the VMC results for $N_{\rm a}=102$ is plotted in the figure,
the analytic expression for the Gutzwiller wave function is available
as in the text.
The non-interacting case is also shown.

\FIG
Comparison of momentum distribution function
between the GWF (solid line) and the exact
diagonalization (open and closed circles) at $J/t=2$.
The analytic expressions for the GWF have been obtained
in Ref.\refto{Voll1,Voll2} and given in Appendix A.
Here we show the results of VMC calculations in 60- and 72-site systems
with $5\times10^4$ samples.
For the diagonalization, the data in $N_{\rm a}=4$, 8, 12
and 16 sites for $n=0.5$ and $N_{\rm a}=8$ and 16 sites for $n=0.75$ are shown.

\FIG
Comparison of spin and charge correlation functions
between the GWF (solid line) and the exact
diagonalization (open and closed circles) at $J/t=2$.
Numerical results are obtained similarly as in Fig.\ 4.

\FIG
Correlation factor $\eta(r)$ for three types of Fermi-liquid-type
wave functions.
Typical values of the parameters are chosen for each case.

\FIG
Variational expectation values of total energy in the supersymmetric
case for (a) the RJWF and for (b) the AJWF.
The value of the GWF is shown by an arrow on the vertical axis.
In this calculation $3\times10^4$ samples are used for a 60-site
system.

\FIG
Expectation values of two energy component ${\cal H}_t$ (solid circle)
and ${\cal H}_J$ (open circle) per site for the RJWF for $n=0.5$.
The value of the GWF ($\zeta=0$) is shown by arrows on the vertical axis.
We use $3\times 10^4$ samples for a 60-site system.

\FIG
Energy expectation values for the RJWF at $J/t=1.0$.
Arrow on the vertical axis is the value of the GWF and that of the
exact diagonalization (extrapolated values for
$N_{\rm a}\rightarrow\infty$).

\FIG
Expectation values of two energy component ${\cal H}_t$ (solid circle)
and ${\cal H}_J$ (open circle) per site for the RJWF and for small-$\zeta$
region.
In this VMC calculation $10^5$ samples are collected for a
72-site system.

\FIG
Discontinuity of momentum distribution $n(k)$ at $k_{\rm F}$ as
a function of $n$.
The optimized Fermi-liquid-type wave function is used for each value
of $n$ and $J/t$.
The size of the symbols represent the relative magnitude of possible
error.
For the cases $n\rightarrow 0$ and $J/t\rightarrow 0$ and for the
AJWF it is not easy to determine the value accurately owing to
the difficulty in optimizing the parameters.

\FIG
Variational expectation values of (a) ${\cal H}_t$ and (b) ${\cal H}_J$
per site for the AJWF.
The arrow on the left vertical axis shows the value of the GWF
($\alpha=0$).
The arrows on the right axis in (b) shows the value of the Heisenberg
chain (Bethe Ansatz) multiplied by $n$.
The sample number is $3\times10^4$ for each point and $N_{\rm a}=72$.

\FIG
Total energy of the AJWF for a couple of values of $J/t$ around
the phase transition.
The same symbol with Fig.\ 12 is used for each value of $\beta$.
(a) corresponds the homogeneous regime, (b) is the case near the
phase transition, and (c) is in the phase-separated regime.
The arrow on the left vertical axis shows the value of the GWF.
The Heisenberg value corrected by the electron density is given
by $-n(J/t)\ln2$:
$-1.8195\cdots$ for $n=0.75$ and $J/t=3.5$.

\FIG
Snapshots of the electron configurations in the VMC sweep for
$N_{\rm a}=60$, $n=0.5$ at $J/t=3.4$ (separated phase).
Each horizontal line represents the one-dimensional system.
An solid (open) circle means an up-(down-)spin electron, and an null
space (horizontal line) an empty site.
In the Monte Carlo sweep, the configuration of the system evolves
vertically.
Before taking these snapshots, 3000 MC steps are discarded
for obtaining the statistical equilibrium.
Sampling interval is determined ($I=20$MCS) so as to make the
acceptance ratio per electron larger than the unit.

\FIG
Real-space charge density correlation function
for the AJWF are shown for various values of $\alpha$.
The value of $\beta$ is fixed at 0.75, which gives minima
approximately around the phase transition.
The solid line shows the values of the completely
separated phase ($\alpha=\infty$).
The system size is 60 sites and $5\times10^4$ samples are used.

\FIG
Real-space correlation function $N_r$ for the nearest neighbor sites
and its derivative $\delta_r$ for $r=N_{\rm a}/2$ are plotted as a
function of $\alpha$.
Arrows on the vertical axes of the left and right represent the values
of $\alpha=0$ (GWF) and $\alpha=\infty$ (completely separated state),
respectively.
The used data are the same ones in Fig.\ 15.

\FIG
Spin correlation function
for the AJWF for various values of $\alpha$.
The solid line represents the value of the Heisenberg chain
obtained by the exact diagonalizations, which is corrected by the
electron density.
The system size is 60 sites and $5\times10^4$ samples are used.

\FIG
Phase diagram of the 1D $t$-$J$ model calculated in the TL-liquid
wave function (5.1).  The curves show the contours of constant correlation
exponent $K_\rho$.
The used system has 100 sites.

\FIG
Comparison of momentum distribution function $n(k)$, spin and charge
correlation functions, $S(k)$ and $N(k)$ among the TL-liquid wave function
(bold solid line), the Fermi-liquid-type wave function (dot) and the exact
diagonalization (open diamond, open circle and exceptionally thin
solid line for $N(k)$ of $J/t=0)$ for (a) $J/t=0$ ($\nu=0.75$
for the TL-liquid state, $\zeta=1.6$ for the RJWF), (b) $J/t=1$ ($\nu=0.37$,
$\zeta=0.7$) and (c) $J/t=2.6$ ($\nu=-0.19$ for the TL-liquid state,
$\alpha=0.4$ and $\beta=1.0$ for the AJWF).
For the variational calculations, $10^5$ samples are averaged for
220-site systems.
Total energy for each case is also given in digits (the last digit
for each include the ambiguity due to the statistical fluctuations).
For exact diagonalization, 52-site system is used in (a) and up to
16-site systems in (b) and (c).

\FIG
Charge susceptibility vs.\ $n$ for some values of $J/t$.
Symbols are the results of the optimized TL-liquid state.
Solid line for $J/t=0$ and 2\refto{KY2} represents the exact
analytic value.
Dashed line is the result for the non-interacting system.
Dotted line for $J/t=1.0$ and 2.5 is a guide to the eyes.
The sizes of the symbols represent the relative magnitude of possible
error.
$50\sim 210$-site systems are used.

\FIG
Chemical potential as a function of $\delta^2$ for $J/t=0$ and 2.
Symbols indicate different system sizes, namely
$N_{\rm a}=200$ (solid diamond), 150 (open circle), 100 (solid circle),
70 (solid square), 50 (open diamond).

\FIG
Spin susceptibility vs. $n$ for some values of $J/t$.
Open circle is the result of the GWF for $J/t=2$ and open diamond is the
result of the optimized TL-liquid state.
Solid line for $J/t=2$ represent the exact analytic
value.\refto{tjBA2,KY2}
Dashed line is the value for the non-interacting system.
Dotted line for $J/t=1.0$ and 2.5 is a guide to the eyes.
Arrow on the right vertical axis shows the exact value for the
Heisenberg antiferromagnet $2/\pi^2$ for $J/t=2$.\refto{Griffiths}
The sizes of the symbols represent the relative magnitude of possible
error.
$50\sim 210$-site systems are used.

\FIG
Total energy under zero field as a function of $m$ calculated with
the TL-liquid wave function.
For the half filling, where the TL-liquid state is reduced to the GWF,
the value in the supersymmetric case is shown for an example.
For $n=1.0$ (0.5), a 102-(100-)site system is used.
$5\times10^4$ samples are averaged.

\FIG
Comparison of magnetization curves for $n=0.5$ and 1.0.
Symbols represent the result of the variational functions.
The optimized TL-liquid wave functions are used except for the
supersymmetric case,
in which the GWF is substituted.
Solid line is the Bethe Ansatz result for
$J/t=2$.\refto{Griffiths,Quaisser}
Dotted line is the result for the non-interacting system.
The result for the supersymmetric long-range $t$-$J$ model is
shown by dashed line.\refto{Kawakami2,YK}
\par

\FIG
(a) Charge and (b) spin susceptibilities calculated with Gutzwiller
Approximation for the Hubbard model as a function of electron
density for some values of $U/U_{\rm c}$.
Inset in (a) represents $\chi_{\rm c}$ vs. $U/U_{\rm c}$ for the
half filling.
The 1D cosine band is assumed.
\par

\bye